\documentclass[a4paper,12pt]{article}
\pdfoutput=1 

\usepackage{jinstpub} 
\usepackage{xspace}
\usepackage[toc,page]{appendix}
\usepackage{array}
\usepackage{tabulary}
\newcolumntype{K}[1]{>{\centering\arraybackslash}p{#1}}

\usepackage{graphicx}
\usepackage{dcolumn}
\usepackage{bm}

\usepackage{graphicx}  
\usepackage{xspace}
\usepackage{units}
\usepackage{comment}
\usepackage{xcolor}

\usepackage{natbib}
\usepackage{hyperref}

\title{Use of Neutrino Scattering Events with Low Hadronic Recoil to Inform Neutrino Flux and Detector Energy Scale}

\newcommand{\Florida}{University of Florida, Department of Physics, Gainesville, FL 32611}

\newcommand{\CBPF}{Centro Brasileiro de Pesquisas F\'{i}sicas, Rua Dr. Xavier Sigaud 150, Urca, Rio de Janeiro, Rio de Janeiro, 22290-180, Brazil}
\newcommand{\PUCP}{Secci\'{o}n F\'{i}sica, Departamento de Ciencias, Pontificia Universidad Cat\'{o}lica del Per\'{u}, Apartado 1761, Lima, Per\'{u}}

\newcommand{\Pittsburgh}{Department of Physics and Astronomy, University of Pittsburgh, Pittsburgh, Pennsylvania 15260, USA}
\newcommand{\Guanajuato}{Campus Le\'{o}n y Campus Guanajuato, Universidad de Guanajuato, Lascurain de Retana No. 5, Colonia Centro, Guanajuato 36000, Guanajuato M\'{e}xico.}

\newcommand{\Tufts}{Physics Department, Tufts University, Medford, Massachusetts 02155, USA}
\newcommand{\WM}{Department of Physics, College of William \& Mary, Williamsburg, Virginia 23187, USA}
\newcommand{\FNAL}{Fermi National Accelerator Laboratory, Batavia, Illinois 60510, USA}

\newcommand{\MCLA}{Massachusetts College of Liberal Arts, 375 Church Street, North Adams, MA 01247}

\newcommand{\UNI}{Facultad de Ciencias, Universidad Nacional de Ingenier\'{i}a, Apartado 31139, Lima, Per\'{u}}
\newcommand{\Rochester}{University of Rochester, Rochester, New York 14627 USA}

\newcommand{\USM}{Departamento de F\'{i}sica, Universidad T\'{e}cnica Federico Santa Mar\'{i}a, Avenida Espa\~{n}a 1680 Casilla 110-V, Valpara\'{i}so, Chile}
\newcommand{\Geneva}{University of Geneva, 1211 Geneva 4, Switzerland}

\newcommand{\OregonState}{Department of Physics, Oregon State University, Corvallis, Oregon 97331, USA}
\newcommand{\oxford}{Oxford University, Department of Physics, Oxford, OX1 3PJ United Kingdom}

\newcommand{\upenn}{Department of Physics and Astronomy, University of Pennsylvania, Philadelphia, PA 19104}
\newcommand{\AMU}{AMU Campus, Aligarh, Uttar Pradesh 202001, India}

\newcommand{\Mohali}{Department of Physical Sciences, IISER Mohali, Knowledge City, SAS Nagar, Mohali - 140306, Punjab, India}

\newcommand{\york}{York University, Department of Physics and Astronomy, Toronto, Ontario, M3J 1P3 Canada}
\newcommand{\Notre}{Department of Physics, University of Notre Dame, Notre Dame, Indiana 46556, USA}
\newcommand{\ICL}{The Blackett Laboratory,  Imperial College London,  London SW7 2BW, United Kingdom}

\newcommand{\nOregonState}{a}
\newcommand{\nFlorida}{b}
\newcommand{\nPittsburgh}{c}
\newcommand{\nWM}{d}
\newcommand{\nAMU}{e}
\newcommand{\nPUCP}{f}
\newcommand{\nRochester}{g}
\newcommand{\nFNAL}{h}
\newcommand{\nGeneva}{i}
\newcommand{\nCBPF}{j}
\newcommand{\nGuanajuato}{k}
\newcommand{\nNotre}{l}
\newcommand{\nTufts}{m}
\newcommand{\nUSM}{n}
\newcommand{\nUMD}{o}
\newcommand{\nyork}{p}
\newcommand{\nMohali}{q}
\newcommand{\nupenn}{r}
\newcommand{\noxford}{s}
\newcommand{\nMCLA}{t}
\newcommand{\nUNI}{u}
\newcommand{\nICL}{v}

\author[\nOregonState]{A.~Bashyal}
\author[\nFlorida]{D.~Rimal}
\author[\nPittsburgh]{B.~Messerly}
\author[\nWM,\nAMU]{Z.~~Ahmad~Dar}
\author[\nAMU]{F.~Akbar}
\author[\nPUCP]{M.~V.~Ascencio}
\author[\nRochester]{A.~Bercellie}
\author[\nFNAL]{M.~Betancourt}
\author[\nRochester]{A.~Bodek}
\author[\nGuanajuato]{J.~L.~Bonilla}
\author[\nGeneva]{A.~Bravar}
\author[\nRochester]{H.~Budd}
\author[\nCBPF]{G.~Caceres}
\author[\nRochester]{T.~Cai}
\author[\nOregonState,\nCBPF]{M.F.~Carneiro}
\author[\nCBPF]{H.~da~Motta}
\author[\nPittsburgh]{S.A.~Dytman}
\author[\nRochester]{G.A.~D\'{i}az}
\author[\nGuanajuato]{J.~Felix}
\author[\nFNAL,\nNotre]{L.~Fields}
\author[\nWM]{A.~Filkins}
\author[\nRochester]{R.~Fine}
\author[\nPUCP]{A.M.~Gago}
\author[\nTufts]{H.~Gallagher}
\author[\nUSM,\nCBPF]{A.~Ghosh}
\author[\nOregonState]{S.M.~Gilligan}
\author[\nUMD]{R.~Gran}
\author[\nyork,\nFNAL]{D.A.~Harris}
\author[\nRochester]{S.~Henry}
\author[\nMohali]{S.~Jena}
\author[\nFNAL]{D.~Jena}
\author[\nRochester]{J.~Kleykamp}
\author[\nWM]{M.~Kordosky}
\author[\nupenn]{D.~Last}
\author[\nCBPF]{A.~Lozano}
\author[\noxford]{X.-G.~Lu}
\author[\nMCLA]{E.~Maher}
\author[\nRochester]{S.~Manly}
\author[\nTufts]{W.A.~Mann}
\author[\nupenn]{C.~Mauger}
\author[\nRochester]{K.S.~McFarland}
\author[\nUSM]{J.~Miller}
\author[\nFNAL]{J.G.~Morf\'{i}n}
\author[\nPittsburgh]{D.~Naples}
\author[\nWM]{J.K.~Nelson}
\author[\nFlorida]{C.~Nguyen}
\author[\nRochester]{A.~Olivier}
\author[\nPittsburgh]{V.~Paolone}
\author[\nFNAL,\nRochester]{G.N.~Perdue}
\author[\nupenn,\nGuanajuato]{M.A.~Ram\'{i}rez}
\author[\nFlorida]{H.~Ray}
\author[\nRochester]{D.~Ruterbories}
\author[\nOregonState]{H.~Schellman}
\author[\nUNI]{C.J.~Solano~Salinas}
\author[\nPittsburgh]{H.~Su}
\author[\nRochester]{M.~Sultana}
\author[\nTufts]{V.S.~Syrotenko}
\author[\nWM,\nGuanajuato]{E.~Valencia}
\author[\nOregonState]{N.H.~Vaughan}
\author[\nICL]{A.V.~Waldron}
\author[\nRochester]{C.~Wret}
\author[\nUSM]{B.~Yaeggy}
\author[\noxford]{K.~Yang}
\author[\nWM]{L.~Zazueta}

\affiliation[\nOregonState]{\OregonState}
\affiliation[\nFlorida]{\Florida}
\affiliation[\nPittsburgh]{\Pittsburgh}
\affiliation[\nWM]{\WM}
\affiliation[\nAMU]{\AMU}
\affiliation[\nPUCP]{\PUCP}
\affiliation[\nRochester]{\Rochester}
\affiliation[\nFNAL]{\FNAL}
\affiliation[\nGuanajuato]{\Guanajuato}
\affiliation[\nGeneva]{\Geneva}
\affiliation[\nCBPF]{\CBPF}
\affiliation[\nNotre]{\Notre}
\affiliation[\nUSM]{\USM}
\affiliation[\nyork]{\york}
\affiliation[\nMohali]{\Mohali}
\affiliation[\nupenn]{\upenn}
\affiliation[\noxford]{\oxford}
\affiliation[\nMCLA]{\MCLA}
\affiliation[\nTufts]{\Tufts}
\affiliation[\nUNI]{\UNI}
\affiliation[\nICL]{\ICL}

\collaboration{The MINER$\nu$A Collaboration}

\abstract{Charged-current neutrino interactions with low hadronic recoil ("low-$\nu$") have a cross-section that is approximately constant versus neutrino energy.  These interactions have been used to measure the shape of neutrino fluxes as a function of neutrino energy at accelerator-based neutrino experiments such as CCFR, NuTeV, MINOS and MINERvA.  In this paper, we demonstrate that low-$\nu$ events can be used to measure parameters of neutrino flux and detector models and that utilization
of event distributions over the upstream detector face can discriminate among parameters
that affect the neutrino flux model.  From fitting a large sample of low-$\nu$ events obtained by exposing MINERvA
to the NuMI medium-energy beam, we find that the best-fit flux parameters are within their {\it a priori} uncertainties, but the energy scale of muons reconstructed in the MINOS detector is shifted by 3.6\% (or 1.8 times the {\it a priori} uncertainty on that parameter).  These fit results are now used in all MINERvA cross-section measurements, and this technique can be applied by other experiments operating at MINERvA energies, such as DUNE.}

\begin{document}
\maketitle
\flushbottom

\section{Introduction}

Precise prediction of the neutrino flux from accelerator-based neutrino beams is a critical ingredient in neutrino physics.  For example, the extraction of neutrino oscillation parameters in long-baseline neutrino experiments requires detailed simulations of reconstructed energy spectra, and neutrino flux predictions are the starting point of these simulations.  Measurements of neutrino interaction cross-sections and other parameters in near detectors rely even more heavily on neutrino flux predictions, as they cannot take advantage of the experimental tuning of the flux model via the near detector used in long-baseline measurements.

The accelerator-based neutrino community has built a toolbox for improving flux predictions and estimating their uncertainties.  This toolbox includes use of external hadron production data~\cite{Aliaga:2016oaz,Abe:2012av} as well as measurements made in neutrino detectors.  The latter is challenging because measurement of neutrino fluxes with a neutrino detector requires a "standard-candle" process with a known neutrino cross-section, and few-GeV neutrino cross-sections are generally poorly known.  Neutrino scattering on electrons, a precisely calculable electroweak process, is one such standard candle, but because the final state electron energy is weakly correlated with the incoming neutrino energy, it constrains the flux normalization but provides little information about the shape of the flux versus energy.  

Charged-current neutrino-nucleus scattering with low hadronic recoil ("low-$\nu$") is another process that has been used as a standard candle.    The inclusive $\nu_\mu$ charged-current cross-section can be expressed as: 

\begin{equation*}
    \frac{\mathrm{d}\sigma}{\mathrm{d}\nu} =
    \frac{G^2_F M}{\pi}
    \int_0^1 \bigg(
        F_2
        - \frac{\nu}{E_\nu}  \left[F_2 + xF_3\right]
        + \frac{\nu}{2E_\nu^2}  \left[\frac{Mx(1-R_L)}{1+R_L}F_2\right] 
        + \frac{\nu^2}{2E_\nu^2}  \left[\frac{F_2}{1+R_L} + xF_3\right]
    \bigg)\,\mathrm{d}x,
    \label{eq:dsigma-dnu}
\end{equation*}
where $E_\nu$ is the neutrino energy, $\nu$ is the energy transferred from the neutrino to the hadronic final state, $x$ is the Bjorken scaling variable, $G_F$ is the Fermi constant, $M$ is the struck nucleon's mass, $F_2$ and $xF_3$ are structure functions, and $R_L$ is the structure function ratio $F_2/(2xF_1)$~\cite{DeVan:2016rkm}.  In the limit that $\nu$/$E_\nu$ is small, all of the energy-dependent terms in the equation above vanish, and the cross-section becomes a constant that is independent of energy.   Although the absolute cross-section for this process is not well known, the fact that it is expected to be independent of neutrino energy means that it can be used to measure the shape of the neutrino flux.   MINOS~\cite{Adamson:2009ju}, and MINERvA~\cite{DeVan:2016rkm,Ren:2017xov}  have used this process to extract the neutrino energy spectrum of the Low Energy (LE) configuration of the Neutrinos at the Main Injector (NuMI) beam~\cite{Adamson:2015dkw}.  

In this paper, we take the low-$\nu$ method a step further and use events with low hadronic energy to identify specific aspects of the flux and detector models that may be inaccurate.  We further use the spatial distribution of low-$\nu$ events across the face of the detector to disentangle various effects.  This technique is applicable to other on-axis neutrino experiments operating at similar energies, and could be exploited in detectors which take data at multiple off-axis locations, such as DUNE-PRISM~\cite{Abi:2020evt}.  The paper is organized as follows.  Section~\ref{sec:minerva} describes the MINERvA detector and simulation; Section~\ref{sec:event_selection} describes the reconstruction of low-$\nu$ events in the MINERvA detector.   Fits to the spectra are described in Section~\ref{sec:fits} and conclusions are presented in Section~\ref{sec:conclusion}.  

\section{MINERvA Experiment and Simulation}
\label{sec:minerva}

The MINERvA detector~\cite{Aliaga:2013uqz} is composed of 208 hexagonal planes of plastic scintillator interspersed with other materials. Each plane contains 127 1.7x3.3 cm triangular scintillator strips, arrayed in one of three directions to facilitate three-dimensional track and shower reconstruction. This study uses muon neutrino interactions in the inner tracker region, which is composed entirely of plastic scintillator planes.  The tracker is surrounded at its outer edges and on the downstream end by scintillator planes separated by 0.2 cm-thick lead sheets, called the electromagnetic calorimeter (ECAL).  Surrounding and downstream of the ECAL is a hadronic calorimeter (HCAL) composed of scintillator interspersed with steel.  The upstream portion of MINERvA contains scintillator interspersed with passive targets made of carbon, iron, lead, water, and helium.  This region was designed for comparing cross-sections across different nuclei and is not used for this study.  The MINERvA detector is positioned 2 m upstream of the magnetized MINOS near detector, which is used to analyze the charge and momentum of muons exiting the back of MINERvA. 

MINERvA is approximately on-axis in the NuMI beamline; the beamline is described in detail in Ref.~\cite{Adamson:2015dkw}.  NuMI begins with a 120 GeV proton beam, which is directed onto a 2-interaction length graphite target.  The target is composed of 48 rectangular fins, each 7.4 (horizontal) x 63 (vertical) x 24 (longitudinal) mm$^3$, and two additional fins rotated by 90 degrees about the beam axis that are used for beam alignment.  The beam hits the target at the horizontal center of the target but is shifted towards the top of the target in the vertical direction, yielding a hadron distribution that is roughly left-right symmetric but with more hadrons exiting the top of the target than the bottom.  The total effective length of the target is 1.2 m.  It is composed of POCO graphite with a density of 1.78 g/cm$^3$.  

Pions and kaons produced in the target are focused using two parabolic focusing horns after which they decay in a 675 m long decay pipe.  The MINERvA detector sits 1032 m downstream of the first focusing horn and is offset from the beam center by -56 cm in the x direction and -53 cm in the y direction where x is left-right and y is top-bottom\footnote{We use the beam coordinate system where the z axis points downstream along the center of the beam, the y axis points upward, and
the x axis is horizontal pointing to beam left.}.
Data for this study were taken between September 9,2013 and February 6,2015. During this period NuMI was configured to focus positively charged particles, resulting in a primarily muon neutrino beam.  NuMI was operated in the ME (Medium Energy) configuration, where the target began 1.43 m upstream of the front face of the first focusing horn and the second horn was 21 m downstream of the first horn. The focusing peak of the muon neutrino flux in this configuration was approximately 6 GeV.    

Simulated MINERvA data begins with a Geant4 simulation of the NuMI beamline.  We use g4numi version v6r3, based on Geant4 version 4.9.3p6 with the FTFP\_BERT physics list.  The beam simulation is corrected with data from hadron production measurements~\cite{Aliaga:2016oaz}.  

Neutrino interactions in the MINERvA detector are simulated using the GENIE~\cite{Andreopoulos:2009rq,Andreopoulos:2015wxa} event generator version 2.12.6.   Within this framework, Quasi-elastic events are simulated using the Llewellyn-Smith formalism~\cite{LlewellynSmith:1971uhs} with BBBA05~\cite{Bradford:2006yz} and a dipole axial form factor with axial mass of 0.99 GeV; resonant pion production uses the Rein-Sehgal model~\cite{REIN198179} with an axial mass of 1.12 GeV; deep inelastic scattering uses the Bodek-Yang model~\cite{Bodek:2004pc}.  The initial state nuclear model uses a Relativistic Fermi Gas~\cite{SMITH1972605} with an additional high momentum tail as prescribed by Bodek and Ritchie~\cite{Bodek:1981wr}.  Final state interactions of hadrons following the initial hard scatter are simulated using the INTRANUKE h-A model~\cite{Dytman:2007zz}.  

MINERvA makes several modifications to the base GENIE model that are collectively known as MINERvA tune v1.  These modifications are as follows:
\begin{itemize}
    \item Low-$Q^2$ quasi-elastic interactions are modified using The Valencia~\cite{Nieves:2004wx} RPA description, as described in ~\cite{Gran:2017psn}.  
    \item Valencia model~\cite{Nieves:2011pp,Gran:2013kda,Schwehr:2016pvn} two-particle, two-hole (2p2h) events are added to the GENIE base model and enhanced using a fit to MINERvA inclusive data~\cite{Rodrigues:2015hik,Ruterbories:2018gub}.  
    \item Non-resonant pion production is suppressed to 40\% of its original strength based on a re-analysis of bubble chamber data~\cite{Rodrigues:2016xjj}.  
\end{itemize}

The response of the MINERvA detector is simulated using Geant4 version 4.9.3p6 with the QGSP\_BERT physics list validated with measurements using a scaled-down version of the detector operated in a hadron test beam~\cite{Aliaga:2015aqe}.  The MINERvA readout and calibration are simulated as described in Ref.~\cite{Aliaga:2013uqz}.  Overlapping events (pile-up) are simulated by overlaying randomly sampled data spills on generated Monte Carlo events, scaled appropriately to simulate different periods of intensity during the running.

 
\section{Low-$\nu$ Event Reconstruction}
\label{sec:event_selection}

The MINERvA detector collects charge depositions (hits) throughout each 10 $\mu$s NuMI spill.  After being read out and calibrated as described in Ref.~\cite{Aliaga:2013uqz}, the hits are correlated in time into so-called time slices.  These are collections of hits consistent with energy depositions from a single neutrino interaction.  Within each time slice, a Kalman filter is used to identify tracks in both the MINERvA and MINOS detectors.  Tracks in the two detectors are then matched based on both time and spatial information.  These matched tracks are deemed to be muons, the only particles capable of producing tracks in both detectors.  To estimate the energy of the non-muon hadronic recoil system, all other hits in MINERvA that are not on the muon track are grouped together and corrected for passive material and neutral particle content using the Monte Carlo simulation described in section~\ref{sec:minerva}.  For the purposes of this study, events are deemed to be low-$\nu$ if the hadronic recoil is less than 800 MeV.  This threshold provides sufficient balance between the size of the sample, which decreases with the threshold, and the energy dependence of the low-$\nu$ cross section, which decreases as the threshold and $\nu/E_\nu$ increase.    

\begin{figure}
    \centering
    \includegraphics[width=0.48\textwidth]{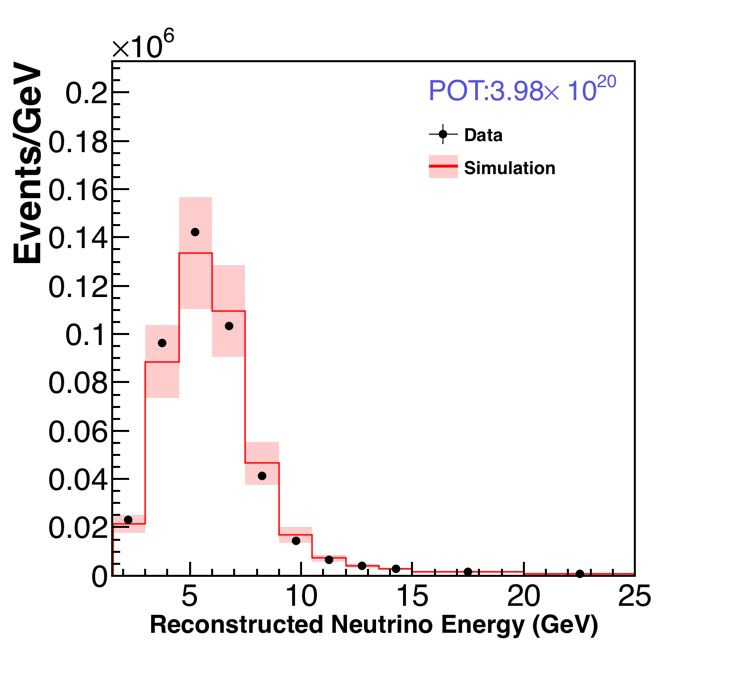}
    \includegraphics[width=0.48\textwidth]{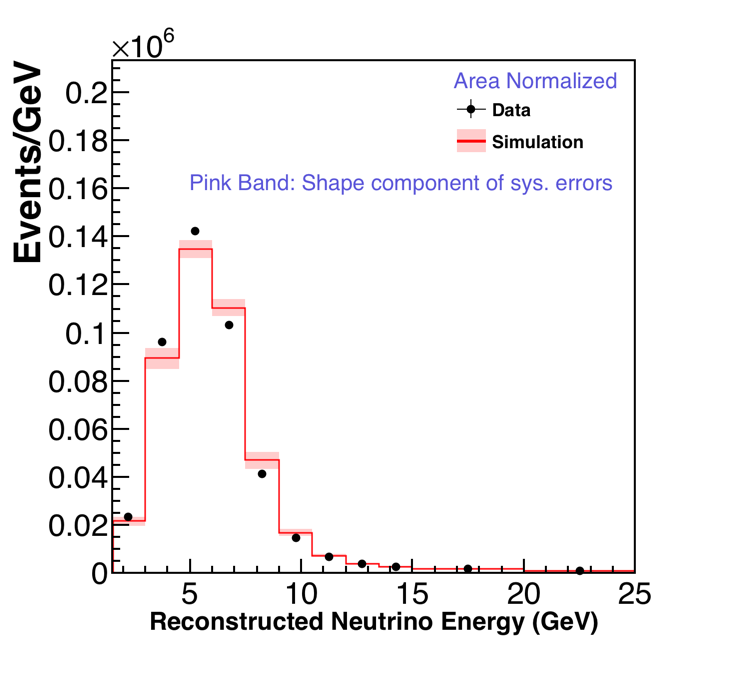}
    \caption{Distribution of reconstructed neutrino energy in low-$\nu$ events in MINERvA data and simulation. Data and simulated low-$\nu$ events with absolute normalization are shown in the left plot. The right plot shows the area normalized data and simulated low-$\nu$ events. The pink band is the systematic uncertainties on simulated events.  These error bands as well as all others shown in this paper represent one standard deviation of uncertainty (equivalent to a 68\% confidence interval).}
    \label{fig:breakdown}
\end{figure}

An estimate of neutrino energy is formed by summing the muon energy and the hadronic recoil energy. The muon energy is derived from the range in MINERvA combined with range in MINOS for muons stopping in MINOS or based on bend in the MINOS magnetic field otherwise.  The distribution of neutrino energy in data and the simulation is shown in figure~\ref{fig:breakdown}, with the simulation both absolutely normalized (left) and area-normalized to the same number of events as data (right).  A significant discrepancy between the data and simulation is apparent.  

The simulated distribution is subject to a number of systematic uncertainties.  All of these have been described in previous MINERvA publications, so we mention them briefly here and include references with more detail on how they are assessed.  
\begin{itemize}
\item Neutrino flux uncertainties, arising from models of hadron production in the target and other beamline materials, as well as accuracy of the simulated beam and focusing system~\cite{Aliaga:2016oaz}.
\item GENIE interaction model uncertainties, arising from both final state and primary interaction models~\cite{Andreopoulos:2009rq,Andreopoulos:2015wxa}. 
\item Additional model-related uncertainties assessed on the MINERvA modifications to GENIE~\cite{Ruterbories:2018gub}.  
\item Uncertainties in the hadronic response of the MINERvA detector~\cite{Mousseau}.  
\item Uncertainties associated with reconstruction of muon tracks in MINERvA and MINOS~\cite{Mousseau}.  
\end{itemize}

  MINERvA assesses systematic uncertainties by varying a parameter in the simulation by its 1-$\sigma$ uncertainty, recomputing the distribution in question, and taking the difference as a systematic uncertainty. In the case of neutrino flux uncertainties, where there are many underlying parameters, many varied distributions are formed by randomly sampling parameters from their probability distributions and taking the RMS of the resulting distributions as the systematic uncertainty. For area-normalized (shape-only) uncertainties, the varied distributions are renormalized to have the same number of events as the original distribution.

\begin{figure}
    \centering
    \includegraphics[width=0.48\textwidth]{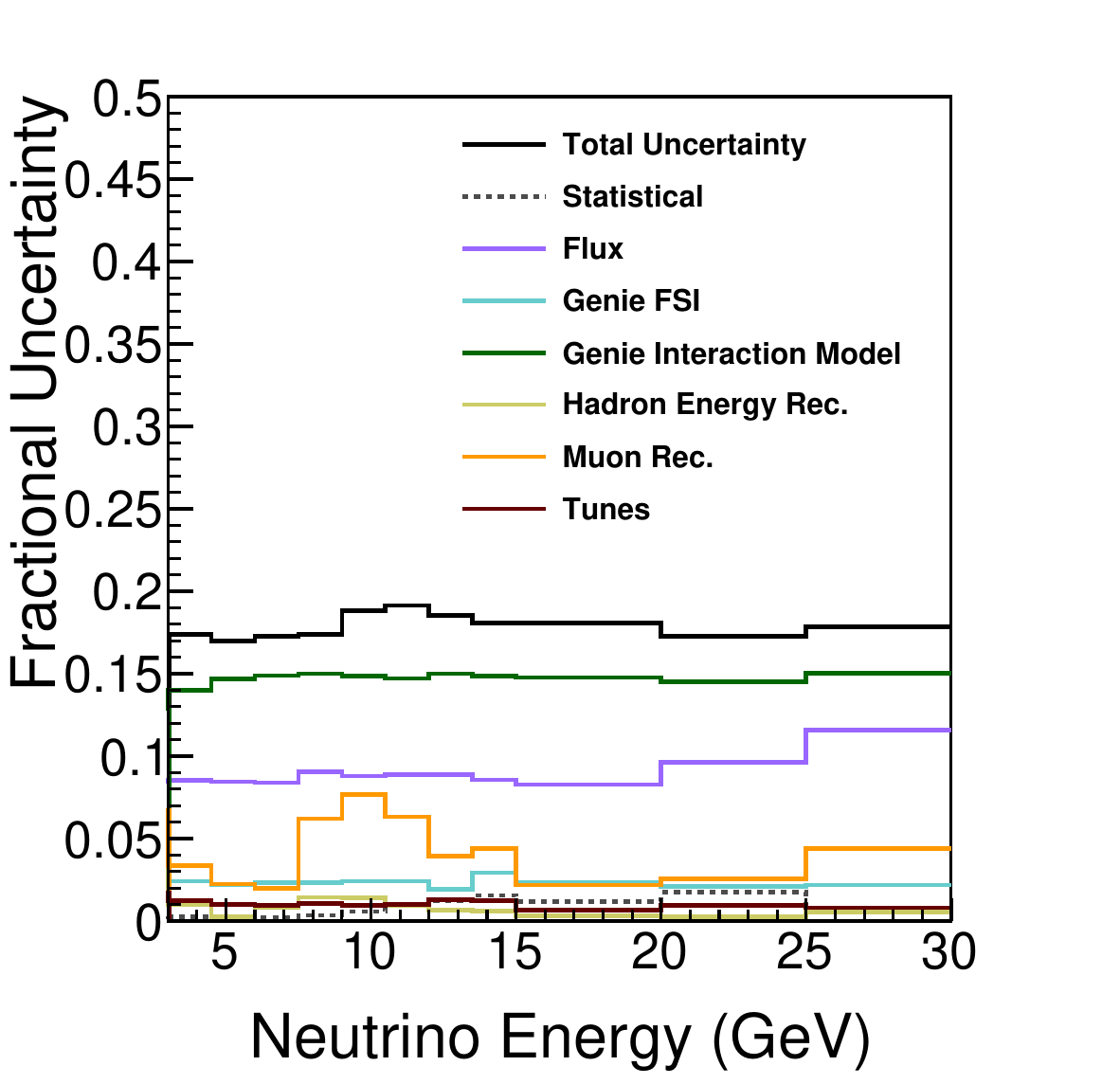}
    \includegraphics[width=0.48\textwidth]{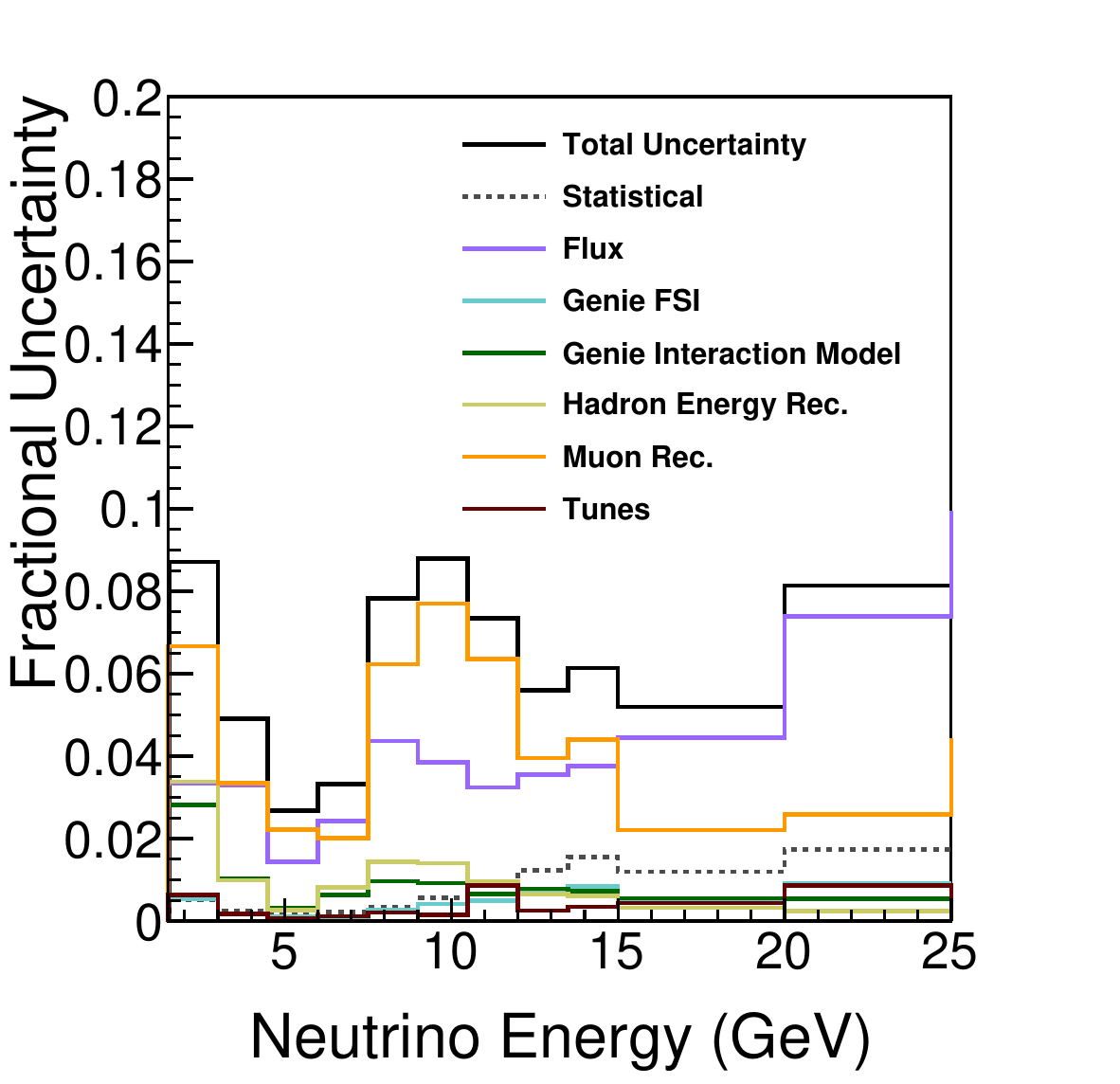}
    \caption{Summary of fractional systematic uncertainties on the simulated neutrino energy distribution for low-$\nu$ events shown in figure~\ref{fig:breakdown}. The left plot shows the total uncertainties and the right plot shows the area-normalized (shape-only) uncertainties.  }
    \label{fig:lownu_systematics}
\end{figure}

A summary of the fractional uncertainty as a function of neutrino energy due to each of the sources enumerated above is shown in figure~\ref{fig:lownu_systematics}.  The largest components of the total uncertainties are associated with the GENIE interaction model and the neutrino flux.  However, the muon reconstruction uncertainty dominates the area-normalized uncertainties at most neutrino energies. This is because shifts in the muon energy scale can shift events in and out of the focusing peak, creating relatively large changes in the shape of the energy spectrum. 
Although the total flux uncertainty is relatively flat (figure~\ref{fig:breakdown} left), subdominant components of the flux uncertainty can affect the shape of the neutrino energy distribution and contribute to the area-normalized uncertainty (figure~\ref{fig:breakdown} right).  These include the focusing uncertainties (see figure~\ref{fig:focusing_uncertainties}) and hadron production uncertainties in the high energy tail, which are weakly correlated with hadron production uncertainties in the focusing peak.  

\begin{figure}
    \centering
    \includegraphics[width=0.48\textwidth]{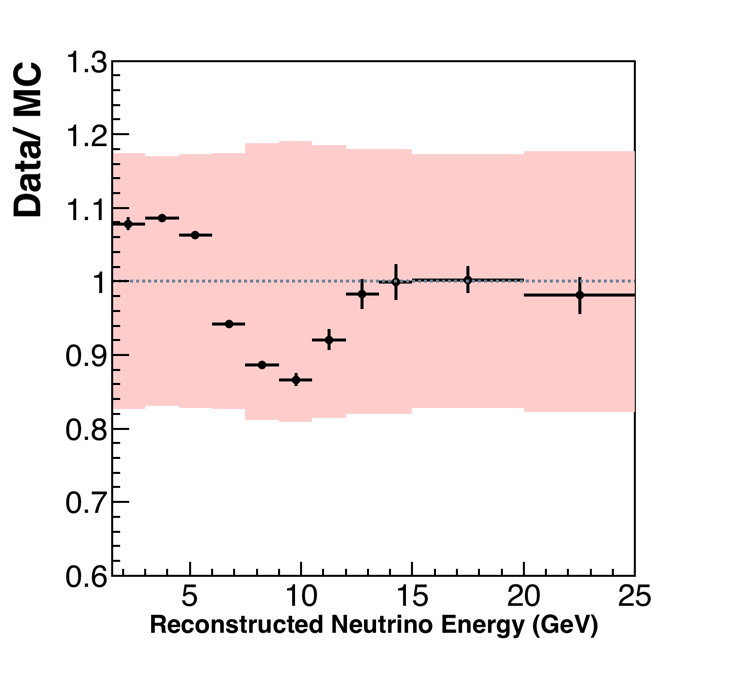}
    \includegraphics[width=0.48\textwidth]{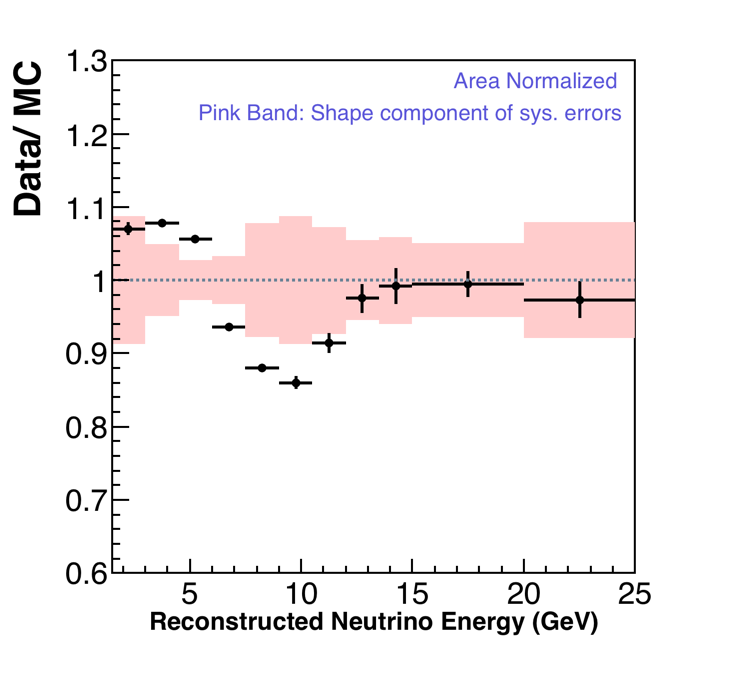}
    \caption{Ratio of the data and simulation for the low-$\nu$ distribution before the fits described in section~\ref{sec:fits}, both absolutely normalized (left) and with data and simulation normalized to the same number of events (right). The pink band is the systematic error band for the simulation, with only the shape component of the systematic uncertainty shown in the right plot.  }
    \label{fig:lownuratiobefore}
\end{figure}

Ratios of data and Monte Carlo are shown in figure~\ref{fig:lownuratiobefore}.  There is a discrepancy between data and simulation that varies substantially as a function of energy.  While this discrepancy is well-covered by the systematic uncertainties, the shape of the discrepancy is much larger than the shape-only component of the systematic uncertainty shown in the right of figure~\ref{fig:lownuratiobefore}.  Most of the sources of systematic uncertainty described above primarily affect the normalization of the low-$\nu$ spectrum but not the shape.  However, there are two sources of uncertainty that could cause discrepancies similar to that shown in figure~\ref{fig:lownuratiobefore}, namely 1) beam focusing parameters and 2) the muon energy scale.  The hadronic energy scale can also modify the shape of the low-$\nu$ spectrum, but because it comprises a very small component of the neutrino energy, it cannot fully account for this discrepancy.  To better understand the source of the discrepancy, fits to the neutrino energy were performed that allowed focusing and muon energy parameters to vary.    


\section{Fits to Energy Spectra}
\label{sec:fits}

Several known sources of uncertainty in MINERvA’s simulation can cause a shift in the
energy spectrum similar to the discrepancy seen in Figure 3. These include the muon energy
scale, which makes up the bulk of the reconstructed neutrino energy in these events.  All muons are assessed a 2\% {\it a priori} uncertainty on the MINOS muon energy scale due to underlying uncertainties in the detector mass and in models of the detector geometry and muon energy loss used in the simulation~\cite{Michael_2008}.  This value was validated with scaled down versions of the MINOS detector~\cite{ADAMSON2006119} and constitutes the total MINOS muon energy uncertainty for muons reconstructed by range.  Samples of muons that can be reconstructed by both range and curvature indicate that there is good agreement between the energy scale of muons reconstructed by range and by curvature.  But MINERvA conservatively adds an additional uncertainty (in quadrature with the 2\% range uncertainty) of 0.6\% (2.5\%) for muons greater than (less than) 1 GeV, based on the precision of the range vs. curvature comparisons~\cite{DiazBautista:2015qyl,Aliaga:2013uqz}.     

\begin{figure}
    \centering
    \includegraphics[width=0.6\textwidth]{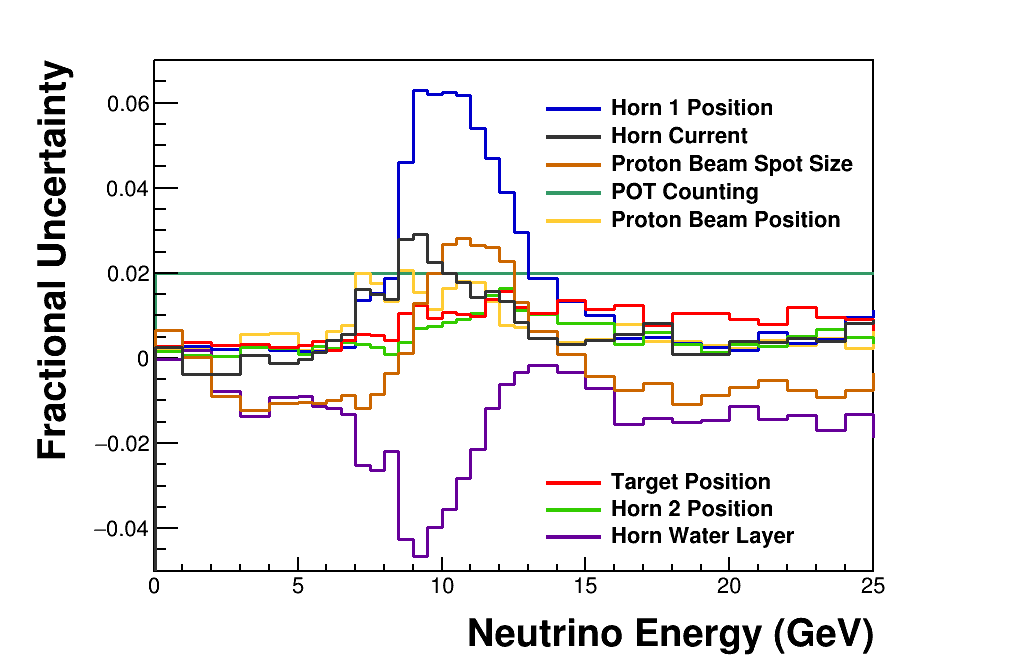}
    \caption{Ratio of predicted neutrino flux with beam parameters shifted by one standard deviation (see table~\ref{tab:focusing_uncertainties}) to the nominal neutrino flux.  The "Beam Position" and "Horn 2 Position" histograms show the combined effect of shifting in x and y.  The "Horn 1 Position" and "Target Position" curves show the combined effect of shifting in x, y and z.  In both of those cases, the effect on the flux of the longitudinal shift (in z) is  small compared to the effects of the transverse shifts (in x and y). }
    \label{fig:focusing_uncertainties}
\end{figure}

Another potential source of the shift is neutrino beam alignment parameters, which preferentially affect high-energy hadrons
that skim the inner edge of the focusing horns and can cause distortions at the falling edge
of the neutrino flux focusing peak.  
  Beam alignment tolerances are shown in table~\ref{tab:focusing_uncertainties}, and the ratio of predicted neutrino flux with these parameters shifted by one standard deviation to the nominal flux is shown in figure~\ref{fig:focusing_uncertainties}.  Shifts of several quantities can create distortions in the predicted neutrino energy spectrum  between 5-15 GeV.  These include the Horn 1 transverse position, the horn current, the size of the horn cooling water layer, and the proton beam position.  However, the shape of the discrepancy within this region and the magnitude per standard deviation vary among the parameters.  

\begin{table}[]
    \centering
    \begin{tabular}{c|c|c|c}
    \hline \hline
        Parameter & Nominal Value & Final 1 $\sigma$ shifts used & Preliminary 1 $\sigma$ shifts  \\
        & & in MINERvA analyses &used in this work \\
         \hline \hline
         Beam Position (X) & 0 mm & 0.4 mm & 1 mm\\
         Beam Position (Y) & 0 mm & 0.4 mm & 1 mm\\
         Beam Spot Size & 1.5 mm  & 0.3 mm & 0.3 mm \\
         Horn Water Layer & 1.0 mm & 0.5 mm & 0.5 mm\\
         Horn Current & 200 kA & 1 kA & 1 kA\\
         Horn 1 Position (X) & 0 mm & 1 mm & 1 mm\\
         Horn 1 Position (Y) & 0 mm & 1 mm & 1 mm\\
         Horn 1 Position (Z) & 30 mm & 2 mm & - \\
         Horn 2 Position (X) & 0 mm & 1 mm & 1 mm\\
         Horn 2 Position (Y) & 0 mm & 1 mm & 1 mm\\
         Target Position (X) & 0 mm & 1 mm & 1 mm \\
         Target Position (Y) & 0 mm & 1 mm & 1 mm\\
         Target Position (Z) & -1433 mm & 1 mm & 3 mm \\
         POT Counting & 0 & 2\% of Total POT & -\\
         Baffle Scraping & 0 & 0.25\% of POT & -\\
         
    \end{tabular}
    \caption{1$\sigma$ tolerances on beam parameters for the NuMI Medium Energy configuration.  The third columns shows the final tolerances provided by NuMI beam experts that are used for MINERvA flux uncertainties and shown in Figure~\ref{fig:focusing_uncertainties}.  The fourth columns show preliminary uncertainties that were used for this work, which began before the final tolerances were known, and are larger than the final tolerances for some parameters.    }
    \label{tab:focusing_uncertainties}
\end{table}

\begin{figure}
    \centering
    \includegraphics[width=0.6\textwidth]{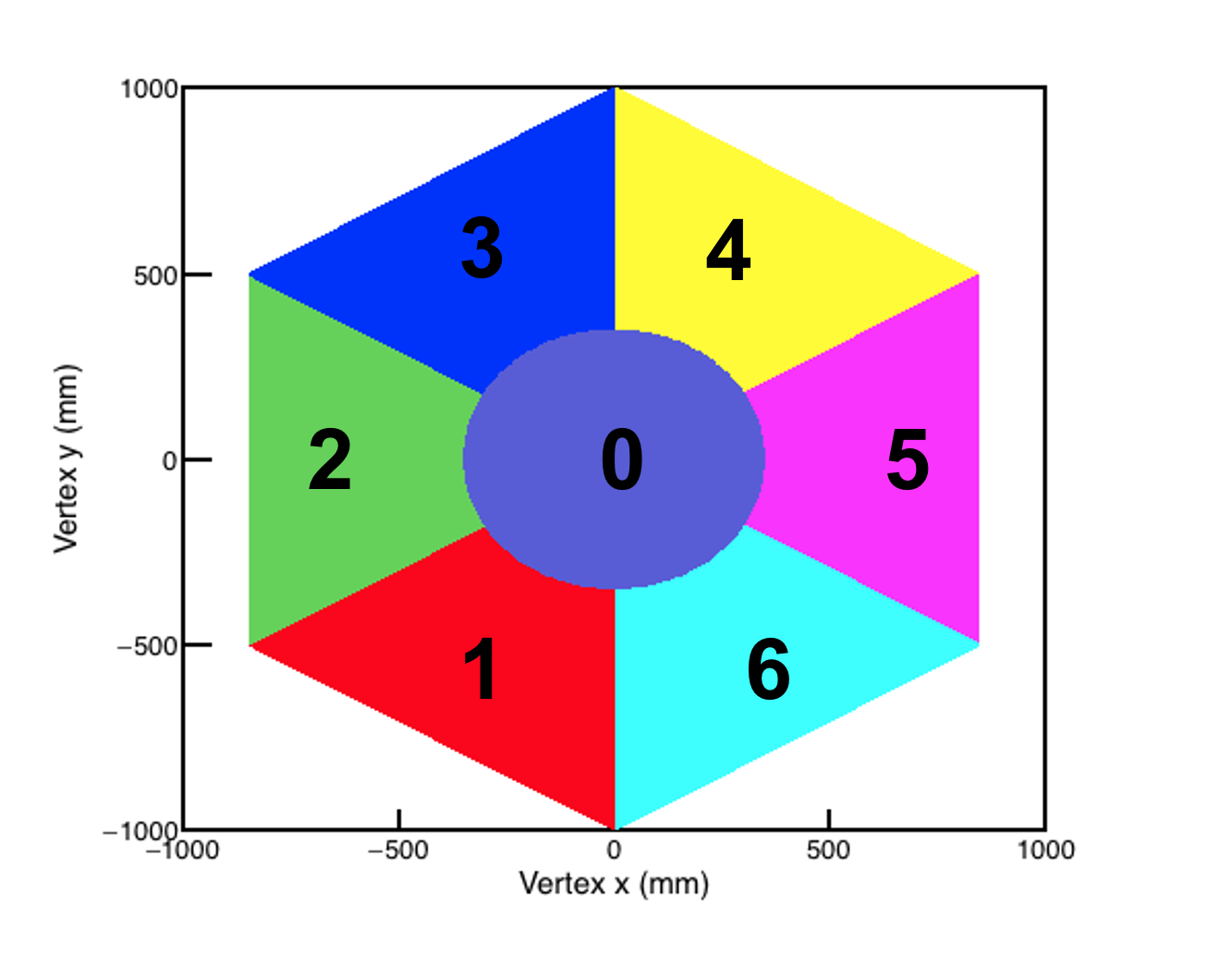}
    \caption{ The seven bins of interaction vertex transverse position used for the fits to low-$\nu$ neutrino energy spectra.      }
    \label{fig:daisy_bins}
\end{figure}

\begin{figure}
    \centering
    \includegraphics[width=0.7\textwidth]{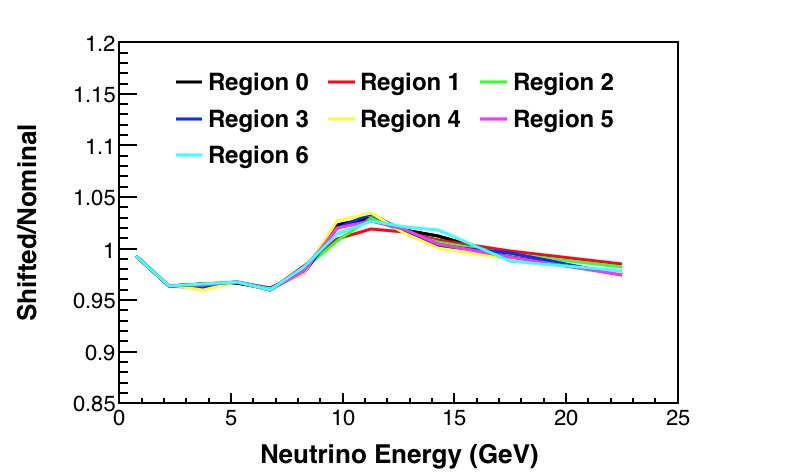}
    \includegraphics[width=0.7\textwidth]{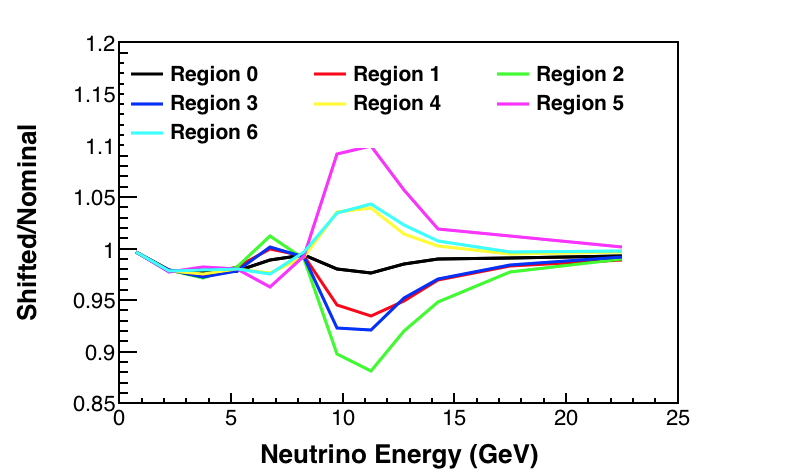}
    \caption{Ratio of varied to nominal neutrino flux for 1 $\sigma$ shifts in the primary proton beam spot size (top) and transverse position on target (bottom), in the seven vertex position bins shown in figure~\ref{fig:daisy_bins}. }
    \label{fig:shifts_daisy_bins}
\end{figure}

The beam focusing parameters can be differentiated by taking advantage of the fact that transverse shifts in beam parameters affect various regions of the detector differently.  To further understand this effect, the low-$\nu$ event sample was separated according to transverse vertex position within the MINERvA detector using the seven bins shown in figure~\ref{fig:daisy_bins}.  The radius of the NuMI beam is larger than the MINERvA detector, so the flux is nearly constant over the face of the detector.  However, the beam is small enough in size that shifts of certain beam parameters from their nominal positions cause variations in flux that are not constant across the detector.  Changes in flux under variations of two alignment parameters are shown in figure~\ref{fig:shifts_daisy_bins}.  In general, transverse shifts to beam components such as the primary proton beam or the horns create different effects in each of the vertex bins, while other types of shifts create a uniform effect in all bins.  

\begin{figure}
    \centering
    \includegraphics[width=0.6\textwidth]{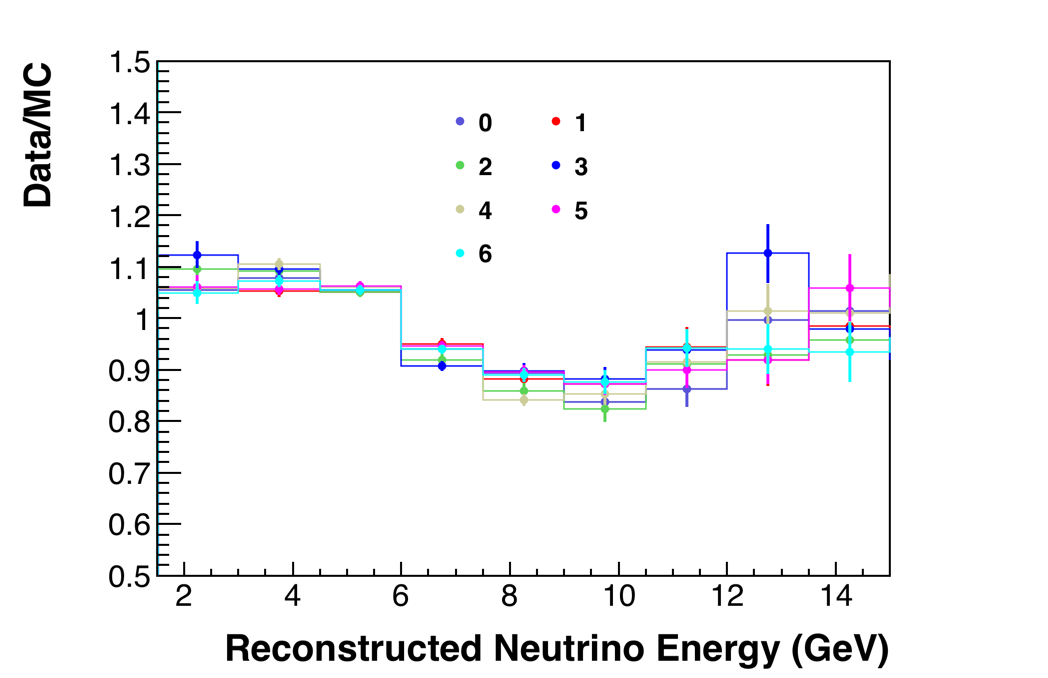}
    \caption{Ratio of low-$\nu$ data to simulation in the seven vertex position bins shown in figure~\ref{fig:daisy_bins}. }
    \label{fig:ratio_bins}
\end{figure}

The ratios of data to simulation of the low-$\nu$ neutrino energy spectra in each of these bins are shown in figure~\ref{fig:ratio_bins}.  The discrepancy is broadly similar in each bin, indicating that the mismodeling is not consistent with a transverse shift of a beam component. 

To further understand which parameters could be the source of the discrepancy, a fit was performed to the low-$\nu$ neutrino energy spectra allowing the beam focusing parameters given in table ~\ref{tab:focusing_uncertainties}\footnote{A parameter for baffle scraping was omitted from the fit, since this parameter has a negligible impact on the predicted neutrino flux} and the MINOS muon energy scale\footnote{The total muon energy combines the muon energy as reconstructed in MINOS with an estimate of energy loss in the MINERvA detector prior to entering MINOS. The MINERvA component of the energy is small, and the fit was found to be insensitive to variations in the energy scale within MINERvA, so only the MINOS component of the energy was varied in the fits. } to vary.  In addition to those parameters that primarily affect the shape of the spectrum, the overall normalization of the spectrum was also allowed to float. The fit minimized a chi squared defined as:

$$\chi ^{2} = \sum _{ij}\frac{(Data' _{ij} - MC'_{ij})^{2} }{\sigma ^{2}_{ij}}$$
where $Data'_{ij}$ ($MC'_{ij}$) is the number of events in the data (simulation) in the $i$th energy bin and the $j$th vertex bin under some set of varied parameters.  The sum is over nine energy bins between 1.5 and 15.0 GeV\footnote{This energy range was chosen because events with energy below 1.5 GeV have very low acceptance into MINOS, while the region above 15 GeV also has relatively few events and is not sensitive to variations of the parameters considered by the fit.}  and the seven vertex bins of figure~\ref{fig:daisy_bins}.  The data and simulated events are reweighted by the flux prediction based on the muon energy scale shift (for data) and focusing parameters (for simulated events). 

The uncertainty is the combined statistical uncertainty of the data and simulation:
$$\sigma_{ij} = \sqrt{\sigma _{Data',ij}^{2} + \sigma _{MC'_{ij}}^{2}}.$$
Fits were performed both with the above $\chi ^2$ and with a modified $\chi^2$ that added a penalty term based on the prior uncertainty on each of the parameters:
$$\chi ^{2}_{prior} = \sum _{ij}\frac{(Data' _{ij} - MC'_{ij})^{2} }{\sigma ^{2}_{ij}} + \sum _{k} (\alpha_{k})^{2},$$
where $\alpha_k$ is the number of standard deviations that parameter k has been shifted from its nominal value.  The standard deviations are taken from the beam parameter tolerances given in the final column of table ~\ref{tab:focusing_uncertainties}.  For the muon energy scale uncertainty, the range uncertainty of 2\% was used as a prior. Alternative versions of the fit were performed adding extra degrees of freedom for the energy scale of muons reconstructed by curvature, but the fit was found to be insensitive to these parameters. 

The result of the fits to the data/simulation ratio is shown in figure~\ref{fig:before_and_after}, while the best fit parameters from the fit are shown in table~\ref{tab:results}, along with their statistical and systematic uncertainties.  Systematic uncertainties on the fit parameters were assessed by shifting sources of uncertainty in the simulation (as enumerated in section~\ref{sec:event_selection} but omitting the beam alignment and muon energy scale parameters that are allowed to vary in fit) by their standard deviations, repeating the fit, and taking the difference between the best fit parameters in the nominal and shifted fits as systematic uncertainties on the fit parameters.  Each source of uncertainty is  shifted one at a time except hadron production flux uncertainties, which are shifted together as described in section~\ref{sec:event_selection}.  All resulting systematic uncertainties are added in quadrature to estimate the total systematic uncertainties quoted in table~\ref{tab:results}.  

\begin{figure}
    \centering
    \includegraphics[width=0.9\textwidth]{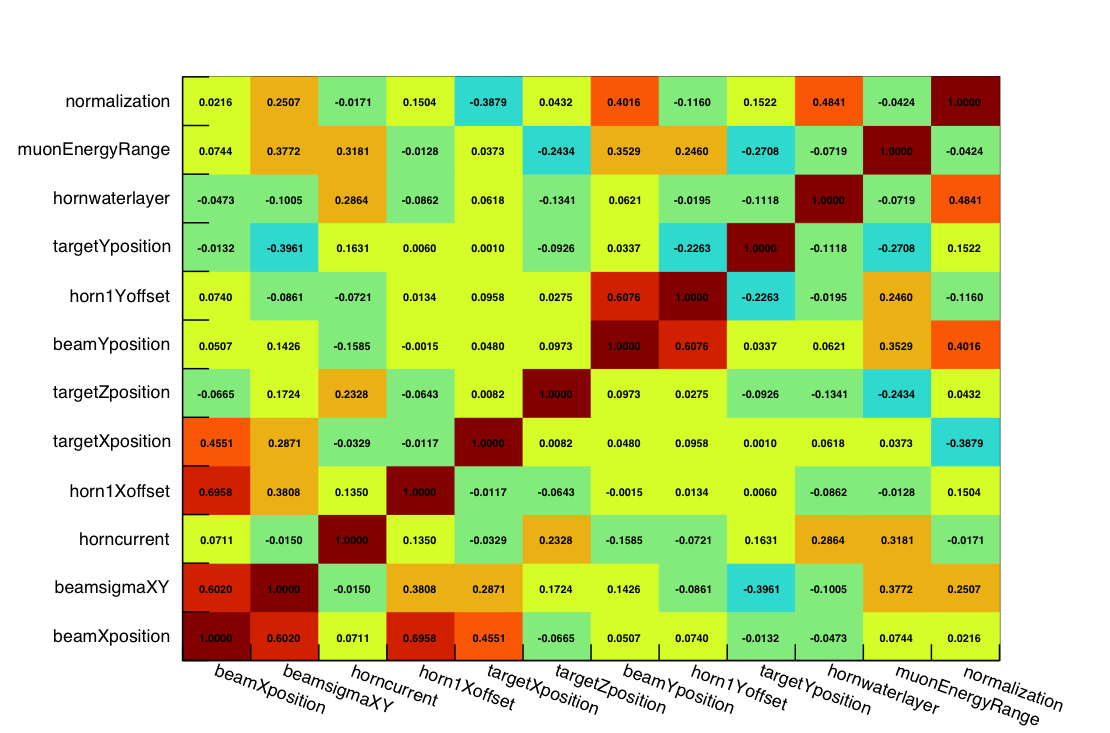}
    \caption{Correlations of best-fit parameters returned by the fit with priors.}
    \label{fig:correlations}
\end{figure}

Because several of the fit parameters can have a similar impact on the low-$\nu$ neutrino energy distributions, there are some correlations between the best fit parameters.  These are shown in Figure~\ref{fig:correlations} for the fit with priors but are similar for the fit without priors.   The largest correlations are between parameters that cause transverse shifts in the same direction (e.g. beam X position and horn 1 x offset), but there are also smaller correlations between several other parameters (e.g. horn current and muon energy scale).  

In both versions of the fit (with and without prior assumptions), the simulation agrees substantially better with the data.  The bulk of the improvement arises from the MINOS muon energy scale, which is shifted by 3.6\% (1.8 times the {\it a priori} standard deviation of this parameter) in the fit with priors.  A few other parameters are pulled by more than one standard deviation from their nominal values, including the target y position in the no-prior fit.  The pulls are more significant in the fit with priors, due in part to the generally smaller post-fit uncertainties.  Parameters with larger pulls in this case include beam y position and target x and y position.  These may be due to the fits reacting to real left/right or up/down asymmetries in the distribution of events across the vertex bins or due to the fact that these parameters are highly correlated with other parameters (see figure~\ref{fig:correlations}).  In any case, the pulls of these parameters have very modest effects on the neutrino energy distribution (3\% changes or less in any neutrino energy bin).  The majority of the change in neutrino energy distribution is provided by the muon energy fit, as shown in the bottom of Figure~\ref{fig:before_and_after}.

\begin{figure}
    \centering
   \includegraphics[width=0.45\textwidth]{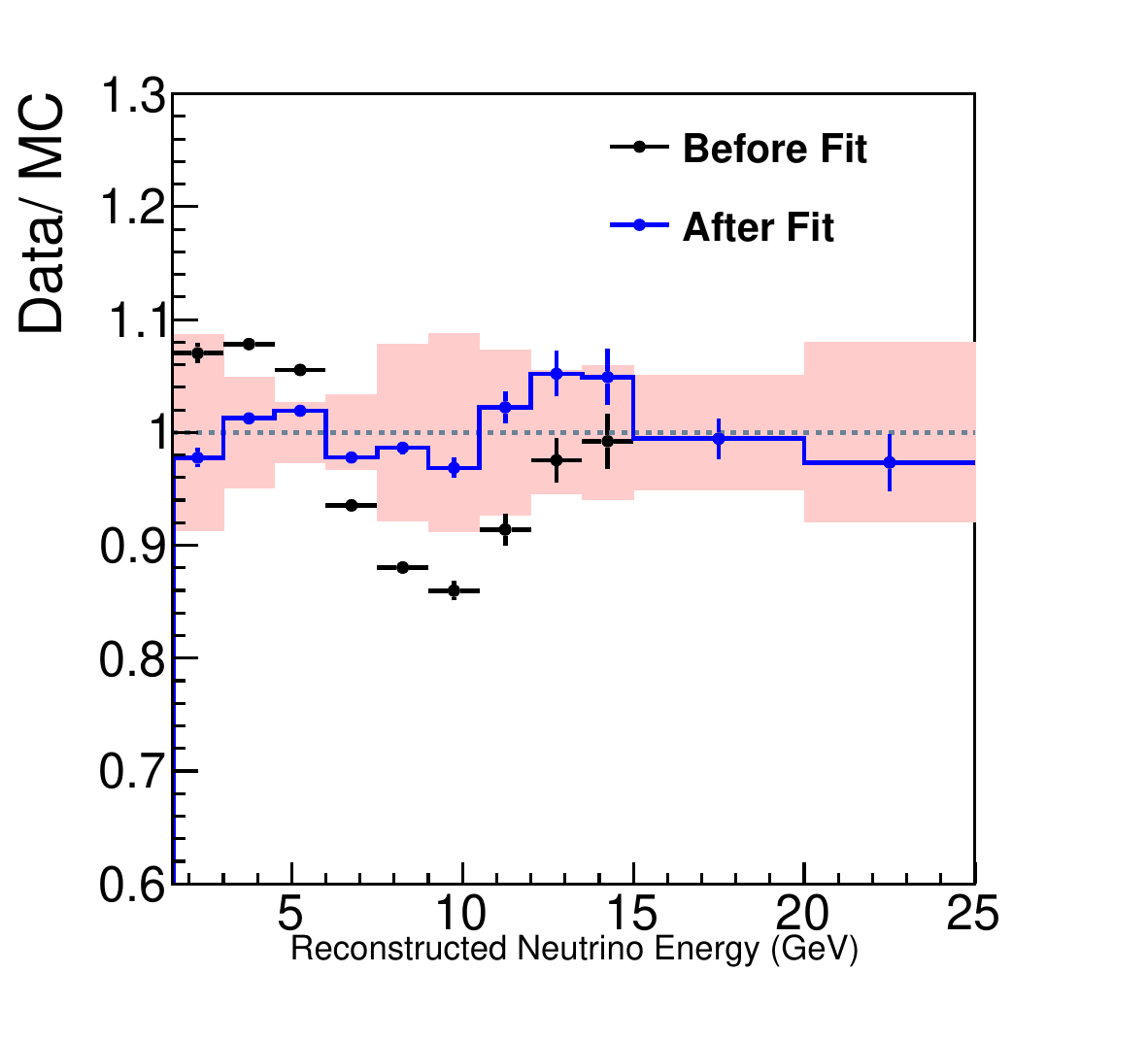}
    \includegraphics[width=0.45\textwidth]{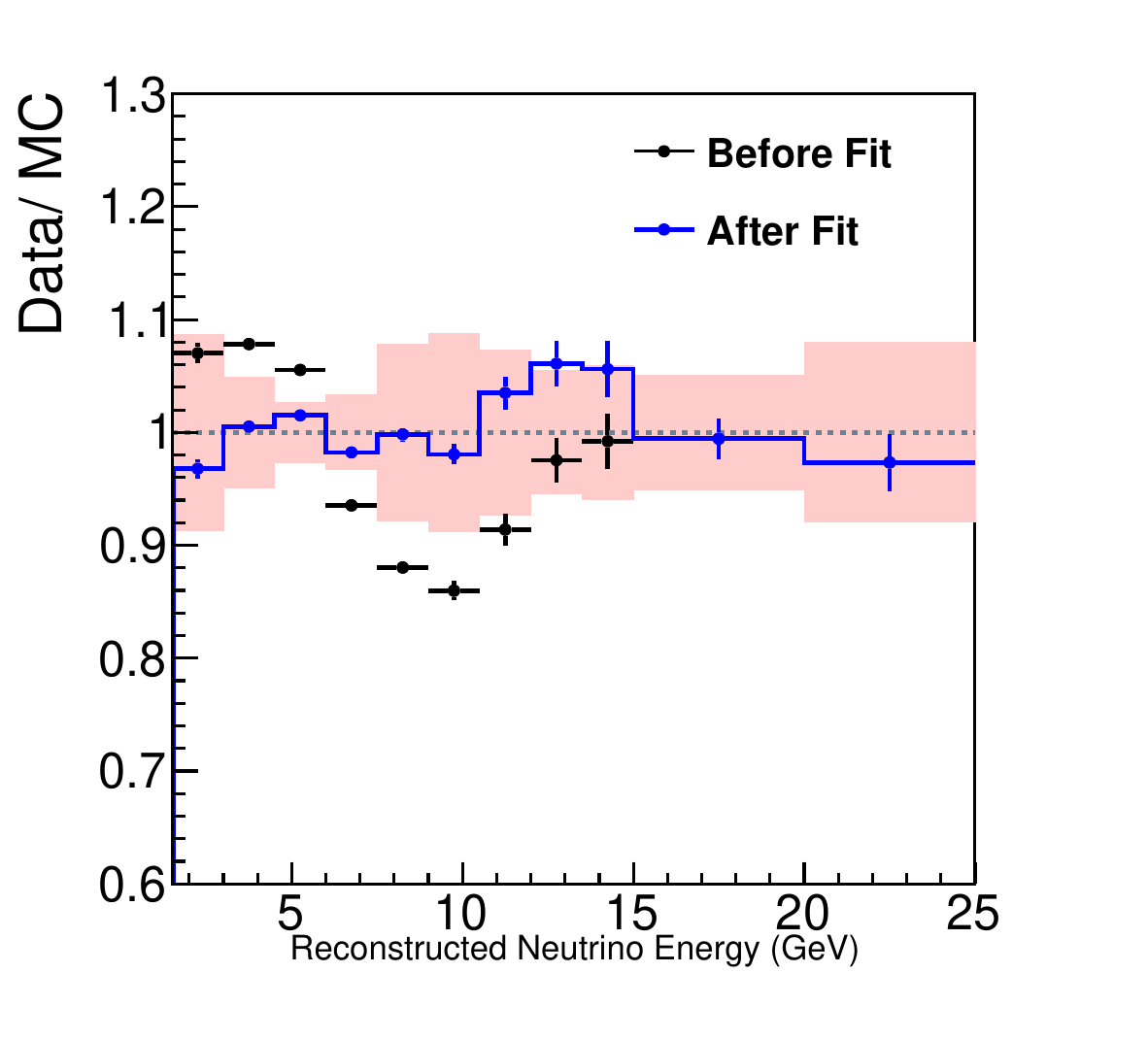}
    \includegraphics[width=0.45\linewidth]{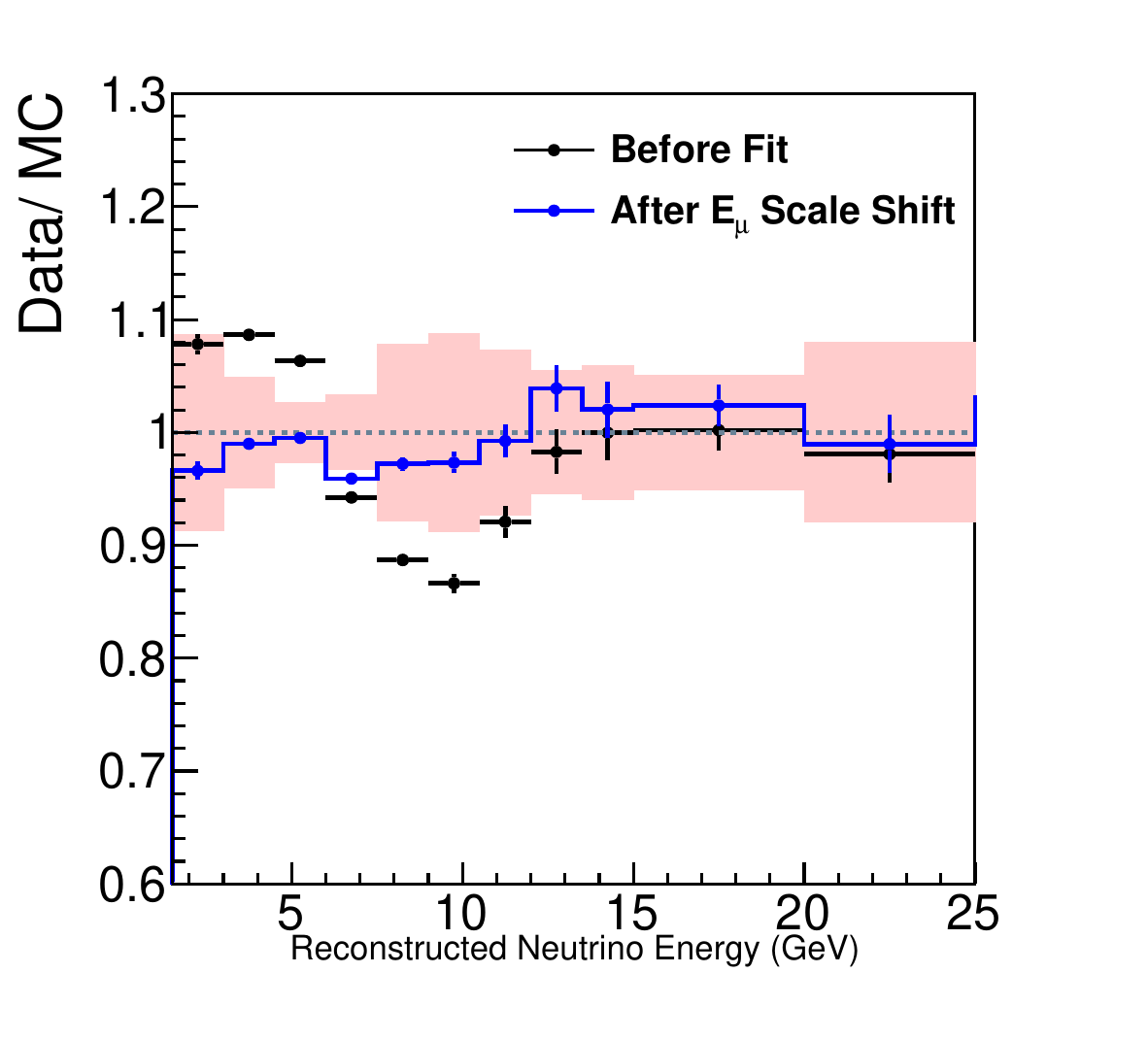}
    \caption{Ratio of low-$\nu$ events in data and simulation before (black) and after (blue) the fits that did not (top left) and did (top right) include a prior penalty term. The bottom plot shows the same events before and after the muon energy scale was shifted by 3.6\% from its nominal value, as prescribed by the fit with priors and as adopted by the MINERvA collaboration.  In all cases, the data and simulation are normalized to the same number of events. The error bars are statistical errors. The pink bands shows the shape component of the systematic uncertainty on the ratio.}
    \label{fig:before_and_after}
\end{figure}

\begin{figure}
    \centering
    \includegraphics[width=0.6\linewidth]{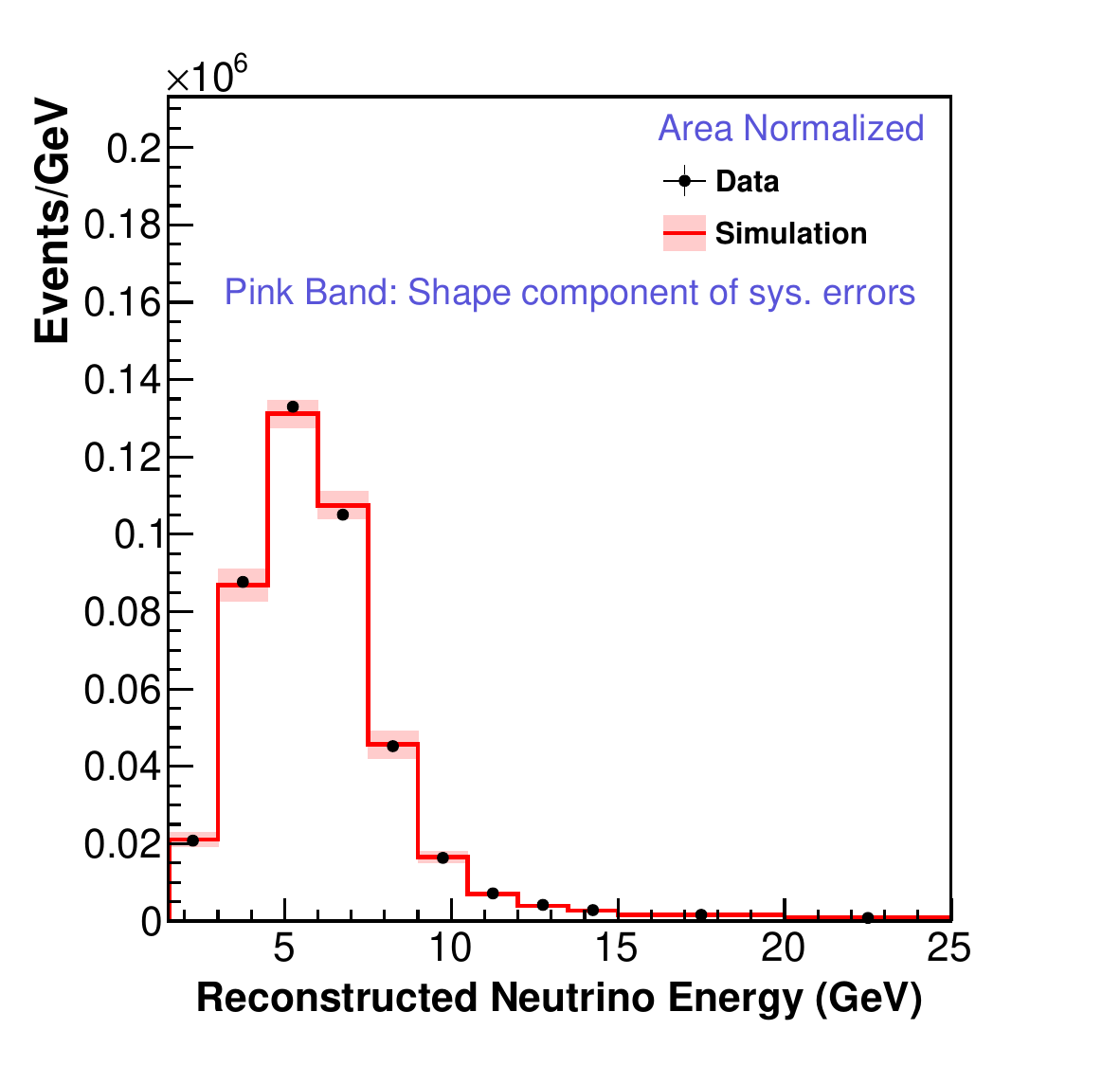}
    
    \caption{Low-$\nu$ distribution after the muon energy scale is shifted in the data. The data and simulation are normalized to the same number of events. The pink band shows the shape component of the systematic uncertainty on the simulated sample.}
    \label{fig:low_nu_distribution_after}
\end{figure}
\begin{table}[]
    \centering
    \begin{tabular}{|c|c|c|c|c|}
    \hline \hline
         Parameter & Nominal & Best Fit  (No Prior) & Best Fit (Prior) \\
         \hline \hline
         Beam Position (X) & 0.0 mm &  $-0.3\pm0.3\pm0.1$ mm & $-0.3\pm0.2\pm0.1$ mm \\
         Beam Position (Y) & 0.0 mm & $0.8\pm0.3\pm0.3$ mm   & $0.7\pm0.2\pm0.2$ mm \\
         Target Position (X) & 0.0 mm &  $-0.8\pm0.3\pm0.1$ mm & $-0.8\pm0.3\pm0.1$ mm \\
         Target Position (Y) & 0.0 mm &  $2.3\pm 0.7\pm1.2$ mm & $1.7\pm0.6\pm0.8$ mm \\
         Target Position (Z) & -1433 mm &  $-1432.4\pm2.4\pm0.3$ mm & $-1431\pm1.8\pm0.3$ mm \\
         Horn 1 Position (X) & 0.0 mm & $-0.3\pm0.4\pm0.5$ mm & $-0.1\pm0.3\pm0.1$ mm \\
         Horn 1 Position (Y) & 0.0 mm & $0.1\pm0.5\pm0.5$ mm & $0.0\pm0.3\pm0.3$ mm \\
         Beam Spot Size & 1.5 mm & $1.41 \pm0.09\pm0.03$ mm & $1.32\pm0.09\pm0.03$ mm \\
         Horn Water Layer & 1.0 mm & $1.2\pm0.3\pm0.05$ mm & $1.3\pm0.25\pm0.1$ mm \\
         Horn Current & 200 kA & $198.0\pm1.4\pm1.4$ kA & $199.1 \pm 0.7\pm0.5$ kA \\
         Muon Energy Scale & 1.0 & $1.032\pm0.004\pm0.008$ & $1.036\pm0.004\pm0.006$ \\
\hline
    \end{tabular}
    \caption{Shift of beam parameters from the fits with and without priors.  }
    \label{tab:results}
\end{table}

To further investigate the fit conclusions, a fit was also performed allowing only beam parameters (and not muon energy scale) to vary.  The results of those fits are available in table ~\ref{tab:flux best fit shifts}.  This alternative fit also achieves good agreement with data and simulation, but results in a shift to the target longitudinal position by 13.6 mm, or more than six times its 2.2 mm tolerance\footnote{The quantity that actually matters here is the relative separation of the horn and target.  The tolerance on that value is the tolerance on the horn 1 Z position (2 mm) added in quadrature with the tolerance on the target Z position (1 mm).}.  NuMI beam experts are confident that the target-horn separation was within this tolerance.  

MINERvA has also used neutrino-electron scattering to constrain the neutrino flux prediction~\cite{Valencia:2019mkf}.  However, that data is primarily sensitive to the normalization of the flux, not the shape, and that data is consistent both with the {\it a priori} flux prediction and with the flux model using all fit results described here.

Since a shift of the MINOS muon energy scale of 1.8 standard deviations is substantially more likely than a shift in the target position of more than 10 standard deviations, we attribute this discrepancy to the MINOS muon energy scale.  For all MINERvA analyses using this data set, the MINOS muon energy scale in the data is shifted by 3.6\%.  This shift is applied to muons reconstructed by both range and curvature because studies have indicated good agreement between these two methods of energy reconstruction and because fits that added extra degrees of freedom for curvature muons did not indicate significant differences between the two samples.  Since the flux predicted by the nominal fit is consistent with the {\it a priori} flux within uncertainties, no correction is made to the flux model.  Because the muon energy is reconstructed primarily using the MINOS near detector, other energy quantities reconstructed in MINERvA events, such as hadronic recoil energy, are presumably not affected, and are not corrected. Figure \ref{fig:low_nu_distribution_after} shows the data and MC after the data has been shifted by 3.6\%.  The ratio of data and MC are shown in the bottom panel of figure~\ref{fig:before_and_after}.  After shifting the energy scale, the residual shape disagreement is within the simulation's 1$\sigma$ uncertainties.         

\begin{table}[]
    \centering
    \begin{tabular}{c|c|c}
    \hline \hline
    Parameter & Nominal Value & New Value \\
    \hline \hline
    Beam Position (X) & 0 mm & -0.2 $\pm$ 0.12 mm\\
    Beam Position (Y) & 0 mm & -0.53 $\pm$ 0.14\\
    Beam Spot Size & 1.5 mm & 1.22 $\pm$ 0.14 mm \\
    Horn Water Layer & 1 mm & 0.895 $\pm$ 0.16 mm \\
    Horn Current & 200 kA & 197.41 $\pm$ 0.76 kA \\
    Horn 1 Position (X) & 0 mm & 0. $\pm$ 0.17 mm \\
    Horn 1 Position (Y)  & 0 mm & -0.39 $\pm$ 0.17 mm \\
    Target Position (X) & 0 mm &  -0.32 $\pm$ 0.17 mm \\
    Target Position (Y) & 0 mm & 1.65 $\pm$ 0.5 mm \\
    Target Position (Z) & -1433 mm & -1419.44 $\pm$ 1.83 mm \\
    \hline
    \end{tabular}
    \caption{Prior and best-fit beam parameters from an alternative fit that did not include the MINOS muon energy scale as a fit parameter. }
    \label{tab:flux best fit shifts}
\end{table}

\section{Conclusion}
\label{sec:conclusion}
The MINERvA collaboration has analyzed a sample of charged current muon neutrino interactions with low hadronic recoil.  A significant discrepancy between data and simulation was observed in the shape of the reconstructed neutrino energy spectrum in this sample.  The discrepancy is consistent with a mismodeling of the alignment parameters of the neutrino beam or of the detector energy scale.  Fits to this data allowing various parameters in the simulation to vary indicate that the discrepancy is most consistent with a 3.6\% shift to the MINOS muon energy scale.  Based on this work, measurements of neutrino cross-sections using this MINERvA dataset include a correction to the MINOS muon energy scale.  This work follows earlier uses of low-$\nu$ samples to measure neutrino flux, but is the first time that this sample has been use to investigate specific sources of neutrino flux and detector mismodeling.  The procedure described here to fit reconstructed low-$\nu$ spectra to flux and detector parameters could be used by other accelerator-based neutrino experiments operating at similar energies.  Additionally, in detectors where the size of the neutrino beam and neutrino detector are similar, the use of transverse vertex position can be used to increase the efficacy of this technique.

\begin{acknowledgments}
This document was prepared by members of the MINERvA Collaboration using the resources of the Fermi National Accelerator Laboratory (Fermilab), a U.S. Department of Energy, Office of Science, HEP User Facility. Fermilab is managed by Fermi Research Alliance, LLC (FRA), acting under Contract No. DE-AC02-07CH11359.
These resources included support for the MINERvA construction project, and support
for construction also
was granted by the United States National Science Foundation under
Award No. PHY-0619727 and by the University of Rochester. Support for
participating scientists was provided by NSF and DOE (USA); by CAPES
and CNPq (Brazil); by CoNaCyT (Mexico); by Proyecto Basal FB 0821, CONICYT PIA ACT1413, and Fondecyt 3170845 and 11130133 (Chile); 
by CONCYTEC (Consejo Nacional de Ciencia, Tecnolog\'ia e Innovaci\'on Tecnol\'ogica), DGI-PUCP (Direcci\'on de Gesti\'on de la Investigaci\'on  - Pontificia Universidad Cat\'olica del Peru), and VRI-UNI (Vice-Rectorate for Research of National University of Engineering) (Peru); NCN Opus Grant No. 2016/21/B/ST2/01092 (Poland); by Science and Technology Facilities Council (UK).  We thank the MINOS Collaboration for use of its near detector data. Finally, we thank the staff of
Fermilab for support of the beam line, the detector, and computing infrastructure.

%
%
%
%
%
%
%
%
%
%
%

\end{acknowledgments}

\bibliographystyle{apsrev4-2}
\bibliography{fluxwiggle}

\begin{thebibliography}{31}%
\makeatletter
\providecommand \@ifxundefined [1]{%
 \@ifx{#1\undefined}
}%
\providecommand \@ifnum [1]{%
 \ifnum #1\expandafter \@firstoftwo
 \else \expandafter \@secondoftwo
 \fi
}%
\providecommand \@ifx [1]{%
 \ifx #1\expandafter \@firstoftwo
 \else \expandafter \@secondoftwo
 \fi
}%
\providecommand \natexlab [1]{#1}%
\providecommand \enquote  [1]{``#1''}%
\providecommand \bibnamefont  [1]{#1}%
\providecommand \bibfnamefont [1]{#1}%
\providecommand \citenamefont [1]{#1}%
\providecommand \href@noop [0]{\@secondoftwo}%
\providecommand \href [0]{\begingroup \@sanitize@url \@href}%
\providecommand \@href[1]{\@@startlink{#1}\@@href}%
\providecommand \@@href[1]{\endgroup#1\@@endlink}%
\providecommand \@sanitize@url [0]{\catcode `\\12\catcode `\$12\catcode
  `\&12\catcode `\#12\catcode `\^12\catcode `\_12\catcode `\%12\relax}%
\providecommand \@@startlink[1]{}%
\providecommand \@@endlink[0]{}%
\providecommand \url  [0]{\begingroup\@sanitize@url \@url }%
\providecommand \@url [1]{\endgroup\@href {#1}{\urlprefix }}%
\providecommand \urlprefix  [0]{URL }%
\providecommand \Eprint [0]{\href }%
\providecommand \doibase [0]{https://doi.org/}%
\providecommand \selectlanguage [0]{\@gobble}%
\providecommand \bibinfo  [0]{\@secondoftwo}%
\providecommand \bibfield  [0]{\@secondoftwo}%
\providecommand \translation [1]{[#1]}%
\providecommand \BibitemOpen [0]{}%
\providecommand \bibitemStop [0]{}%
\providecommand \bibitemNoStop [0]{.\EOS\space}%
\providecommand \EOS [0]{\spacefactor3000\relax}%
\providecommand \BibitemShut  [1]{\csname bibitem#1\endcsname}%
\let\auto@bib@innerbib\@empty
\bibitem [{\citenamefont {Aliaga}\ \emph {et~al.}(2016)\citenamefont {Aliaga}
  \emph {et~al.}}]{Aliaga:2016oaz}%
  \BibitemOpen
  \bibfield  {author} {\bibinfo {author} {\bibfnamefont {L.}~\bibnamefont
  {Aliaga}} \emph {et~al.} (\bibinfo {collaboration} {MINERvA Collaboration}),\
  }\href {https://doi.org/10.1103/PhysRevD.94.092005} {\bibfield  {journal}
  {\bibinfo  {journal} {Phys. Rev. D}\ }\textbf {\bibinfo {volume} {94}},\
  \bibinfo {pages} {092005} (\bibinfo {year} {2016})},\ \bibinfo {note}
  {[Erratum: \href{https://doi.org/10.1103/PhysRevD.95.039903}{Phys. Rev. D
  {\bf 95}, no.3, 039903 (2017)}]},\ \Eprint {https://arxiv.org/abs/1607.00704}
  {arXiv:1607.00704} \BibitemShut {NoStop}%
\bibitem [{\citenamefont {Abe}\ \emph {et~al.}(2013)\citenamefont {Abe} \emph
  {et~al.}}]{Abe:2012av}%
  \BibitemOpen
  \bibfield  {author} {\bibinfo {author} {\bibfnamefont {K.}~\bibnamefont
  {Abe}} \emph {et~al.} (\bibinfo {collaboration} {T2K}),\ }\href
  {https://doi.org/10.1103/PhysRevD.87.012001} {\bibfield  {journal} {\bibinfo
  {journal} {Phys. Rev. D}\ }\textbf {\bibinfo {volume} {87}},\ \bibinfo
  {pages} {012001} (\bibinfo {year} {2013})},\ \bibinfo {note} {[Erratum:
  \href{https://doi.org/10.1103/PhysRevD.87.019902}{Phys. Rev. D {\bf 87}, no.
  1, 019902 (2013)}]},\ \Eprint {https://arxiv.org/abs/1211.0469}
  {arXiv:1211.0469 [hep-ex]} \BibitemShut {NoStop}%
\bibitem [{\citenamefont {Devan}\ \emph {et~al.}(2016)\citenamefont {Devan}
  \emph {et~al.}}]{DeVan:2016rkm}%
  \BibitemOpen
  \bibfield  {author} {\bibinfo {author} {\bibfnamefont {J.}~\bibnamefont
  {Devan}} \emph {et~al.} (\bibinfo {collaboration} {MINERvA}),\ }\href
  {https://doi.org/10.1103/PhysRevD.94.112007} {\bibfield  {journal} {\bibinfo
  {journal} {Phys. Rev. D}\ }\textbf {\bibinfo {volume} {94}},\ \bibinfo
  {pages} {112007} (\bibinfo {year} {2016})},\ \Eprint
  {https://arxiv.org/abs/1610.04746} {arXiv:1610.04746 [hep-ex]} \BibitemShut
  {NoStop}%
\bibitem [{\citenamefont {Adamson}\ \emph {et~al.}(2010)\citenamefont {Adamson}
  \emph {et~al.}}]{Adamson:2009ju}%
  \BibitemOpen
  \bibfield  {author} {\bibinfo {author} {\bibfnamefont {P.}~\bibnamefont
  {Adamson}} \emph {et~al.} (\bibinfo {collaboration} {MINOS}),\ }\href
  {https://doi.org/10.1103/PhysRevD.81.072002} {\bibfield  {journal} {\bibinfo
  {journal} {Phys. Rev. D}\ }\textbf {\bibinfo {volume} {81}},\ \bibinfo
  {pages} {072002} (\bibinfo {year} {2010})},\ \Eprint
  {https://arxiv.org/abs/0910.2201} {arXiv:0910.2201 [hep-ex]} \BibitemShut
  {NoStop}%
\bibitem [{\citenamefont {Ren}\ \emph {et~al.}(2017)\citenamefont {Ren} \emph
  {et~al.}}]{Ren:2017xov}%
  \BibitemOpen
  \bibfield  {author} {\bibinfo {author} {\bibfnamefont {L.}~\bibnamefont
  {Ren}} \emph {et~al.} (\bibinfo {collaboration} {MINERvA}),\ }\href
  {https://doi.org/10.1103/PhysRevD.95.072009} {\bibfield  {journal} {\bibinfo
  {journal} {Phys. Rev. D}\ }\textbf {\bibinfo {volume} {95}},\ \bibinfo
  {pages} {072009} (\bibinfo {year} {2017})},\ \bibinfo {note} {[Erratum:
  \href{https://doi.org/10.1103/PhysRevD.97.019902}{Phys. Rev.D {\bf 97}, no.
  1, 019902 (2018)}]},\ \Eprint {https://arxiv.org/abs/1701.04857}
  {arXiv:1701.04857 [hep-ex]} \BibitemShut {NoStop}%
\bibitem [{\citenamefont {Adamson}\ \emph {et~al.}(2016)\citenamefont {Adamson}
  \emph {et~al.}}]{Adamson:2015dkw}%
  \BibitemOpen
  \bibfield  {author} {\bibinfo {author} {\bibfnamefont {P.}~\bibnamefont
  {Adamson}} \emph {et~al.},\ }\href
  {https://doi.org/10.1016/j.nima.2015.08.063} {\bibfield  {journal} {\bibinfo
  {journal} {Nucl. Instrum. Meth. A}\ }\textbf {\bibinfo {volume} {806}},\
  \bibinfo {pages} {279} (\bibinfo {year} {2016})},\ \Eprint
  {https://arxiv.org/abs/1507.06690} {arXiv:1507.06690} \BibitemShut {NoStop}%
\bibitem [{\citenamefont {Abi}\ \emph {et~al.}(2020)\citenamefont {Abi} \emph
  {et~al.}}]{Abi:2020evt}%
  \BibitemOpen
  \bibfield  {author} {\bibinfo {author} {\bibfnamefont {B.}~\bibnamefont
  {Abi}} \emph {et~al.} (\bibinfo {collaboration} {DUNE}),\ }\href@noop {}
  {\bibinfo {title} {{Deep Underground Neutrino Experiment (DUNE), Far Detector
  Technical Design Report, Volume II DUNE Physics}}} (\bibinfo {year} {2020}),\
  \Eprint {https://arxiv.org/abs/2002.03005} {arXiv:2002.03005 [hep-ex]}
  \BibitemShut {NoStop}%
\bibitem [{\citenamefont {Aliaga}\ \emph {et~al.}(2014)\citenamefont {Aliaga}
  \emph {et~al.}}]{Aliaga:2013uqz}%
  \BibitemOpen
  \bibfield  {author} {\bibinfo {author} {\bibfnamefont {L.}~\bibnamefont
  {Aliaga}} \emph {et~al.} (\bibinfo {collaboration} {MINERvA Collaboration}),\
  }\href {https://doi.org/10.1016/j.nima.2013.12.053} {\bibfield  {journal}
  {\bibinfo  {journal} {Nucl. Instrum. Meth. A}\ }\textbf {\bibinfo {volume}
  {743}},\ \bibinfo {pages} {130} (\bibinfo {year} {2014})},\ \Eprint
  {https://arxiv.org/abs/1305.5199} {arXiv:1305.5199} \BibitemShut {NoStop}%
\bibitem [{\citenamefont {Andreopoulos}\ \emph {et~al.}(2010)\citenamefont
  {Andreopoulos} \emph {et~al.}}]{Andreopoulos:2009rq}%
  \BibitemOpen
  \bibfield  {author} {\bibinfo {author} {\bibfnamefont {C.}~\bibnamefont
  {Andreopoulos}} \emph {et~al.} (\bibinfo {collaboration} {GENIE
  Collaboration}),\ }\href@noop {} {\bibfield  {journal} {\bibinfo  {journal}
  {Nucl. Instrum. Meth. A}\ }\textbf {\bibinfo {volume} {614}},\ \bibinfo
  {pages} {87} (\bibinfo {year} {2010})},\ \Eprint
  {https://arxiv.org/abs/0905.2517} {arXiv:0905.2517 [hep-ph]} \BibitemShut
  {NoStop}%
\bibitem [{\citenamefont {Andreopoulos}\ \emph {et~al.}(2015)\citenamefont
  {Andreopoulos}, \citenamefont {Barry}, \citenamefont {Dytman}, \citenamefont
  {Gallagher}, \citenamefont {Golan}, \citenamefont {Hatcher}, \citenamefont
  {Perdue},\ and\ \citenamefont {Yarba}}]{Andreopoulos:2015wxa}%
  \BibitemOpen
  \bibfield  {author} {\bibinfo {author} {\bibfnamefont {C.}~\bibnamefont
  {Andreopoulos}}, \bibinfo {author} {\bibfnamefont {C.}~\bibnamefont {Barry}},
  \bibinfo {author} {\bibfnamefont {S.}~\bibnamefont {Dytman}}, \bibinfo
  {author} {\bibfnamefont {H.}~\bibnamefont {Gallagher}}, \bibinfo {author}
  {\bibfnamefont {T.}~\bibnamefont {Golan}}, \bibinfo {author} {\bibfnamefont
  {R.}~\bibnamefont {Hatcher}}, \bibinfo {author} {\bibfnamefont
  {G.}~\bibnamefont {Perdue}},\ and\ \bibinfo {author} {\bibfnamefont
  {J.}~\bibnamefont {Yarba}},\ }\href@noop {} {\  (\bibinfo {year} {2015})},\
  \Eprint {https://arxiv.org/abs/1510.05494} {arXiv:1510.05494} \BibitemShut
  {NoStop}%
\bibitem [{\citenamefont {Llewellyn~Smith}(1972)}]{LlewellynSmith:1971uhs}%
  \BibitemOpen
  \bibfield  {author} {\bibinfo {author} {\bibfnamefont {C.~H.}\ \bibnamefont
  {Llewellyn~Smith}},\ }\bibfield  {booktitle} {\emph {\bibinfo {booktitle}
  {{Gauge Theories and Neutrino Physics, Jacob, 1978:0175}}},\ }\href
  {https://doi.org/10.1016/0370-1573(72)90010-5} {\bibfield  {journal}
  {\bibinfo  {journal} {Phys. Rept.}\ }\textbf {\bibinfo {volume} {3}},\
  \bibinfo {pages} {261} (\bibinfo {year} {1972})}\BibitemShut {NoStop}%
\bibitem [{\citenamefont {Bradford}\ \emph {et~al.}(2006)\citenamefont
  {Bradford}, \citenamefont {Bodek}, \citenamefont {Budd},\ and\ \citenamefont
  {Arrington}}]{Bradford:2006yz}%
  \BibitemOpen
  \bibfield  {author} {\bibinfo {author} {\bibfnamefont {R.}~\bibnamefont
  {Bradford}}, \bibinfo {author} {\bibfnamefont {A.}~\bibnamefont {Bodek}},
  \bibinfo {author} {\bibfnamefont {H.~S.}\ \bibnamefont {Budd}},\ and\
  \bibinfo {author} {\bibfnamefont {J.}~\bibnamefont {Arrington}},\ }\href
  {https://doi.org/10.1016/j.nuclphysbps.2006.08.028} {\bibfield  {journal}
  {\bibinfo  {journal} {Nucl.~Phys.~Proc.~Suppl.}\ }\textbf {\bibinfo {volume}
  {159}},\ \bibinfo {pages} {127} (\bibinfo {year} {2006})},\ \Eprint
  {https://arxiv.org/abs/hep-ex/0602017} {arXiv:hep-ex/0602017 [hep-ex]}
  \BibitemShut {NoStop}%
\bibitem [{\citenamefont {Rein}\ and\ \citenamefont
  {Sehgal}(1981)}]{REIN198179}%
  \BibitemOpen
  \bibfield  {author} {\bibinfo {author} {\bibfnamefont {D.}~\bibnamefont
  {Rein}}\ and\ \bibinfo {author} {\bibfnamefont {L.~M.}\ \bibnamefont
  {Sehgal}},\ }\href {https://doi.org/10.1016/0003-4916(81)90242-6} {\bibfield
  {journal} {\bibinfo  {journal} {Annals of Physics}\ }\textbf {\bibinfo
  {volume} {133}},\ \bibinfo {pages} {79 } (\bibinfo {year}
  {1981})}\BibitemShut {NoStop}%
\bibitem [{\citenamefont {Bodek}\ \emph {et~al.}(2005)\citenamefont {Bodek},
  \citenamefont {Park},\ and\ \citenamefont {Yang}}]{Bodek:2004pc}%
  \BibitemOpen
  \bibfield  {author} {\bibinfo {author} {\bibfnamefont {A.}~\bibnamefont
  {Bodek}}, \bibinfo {author} {\bibfnamefont {I.}~\bibnamefont {Park}},\ and\
  \bibinfo {author} {\bibfnamefont {U.-K.}\ \bibnamefont {Yang}},\ }\href
  {https://doi.org/10.1016/j.nuclphysbps.2004.11.208} {\bibfield  {journal}
  {\bibinfo  {journal} {Nucl.~Phys.~Proc.~Suppl.}\ }\textbf {\bibinfo {volume}
  {139}},\ \bibinfo {pages} {113} (\bibinfo {year} {2005})},\ \Eprint
  {https://arxiv.org/abs/hep-ph/0411202} {arXiv:hep-ph/0411202 [hep-ph]}
  \BibitemShut {NoStop}%
\bibitem [{\citenamefont {and E.J.~Moniz}(1972)}]{SMITH1972605}%
  \BibitemOpen
  \bibfield  {author} {\bibinfo {author} {\bibfnamefont {R.~S.}\ \bibnamefont
  {and E.J.~Moniz}},\ }\href {https://doi.org/10.1016/0550-3213(72)90040-5}
  {\bibfield  {journal} {\bibinfo  {journal} {Nucl. Phys. B}\ }\textbf
  {\bibinfo {volume} {43}},\ \bibinfo {pages} {605 } (\bibinfo {year}
  {1972})}\BibitemShut {NoStop}%
\bibitem [{\citenamefont {Bodek}\ and\ \citenamefont
  {Ritchie}(1981)}]{Bodek:1981wr}%
  \BibitemOpen
  \bibfield  {author} {\bibinfo {author} {\bibfnamefont {A.}~\bibnamefont
  {Bodek}}\ and\ \bibinfo {author} {\bibfnamefont {J.~L.}\ \bibnamefont
  {Ritchie}},\ }\href {https://doi.org/10.1103/PhysRevD.24.1400} {\bibfield
  {journal} {\bibinfo  {journal} {Phys.~Rev.~D}\ }\textbf {\bibinfo {volume}
  {24}},\ \bibinfo {pages} {1400} (\bibinfo {year} {1981})}\BibitemShut
  {NoStop}%
\bibitem [{\citenamefont {Dytman}(2007)}]{Dytman:2007zz}%
  \BibitemOpen
  \bibfield  {author} {\bibinfo {author} {\bibfnamefont {S.}~\bibnamefont
  {Dytman}},\ }\href {https://doi.org/10.1063/1.2720468} {\bibfield  {journal}
  {\bibinfo  {journal} {AIP Conf. Proc.}\ }\textbf {\bibinfo {volume} {896}},\
  \bibinfo {pages} {178} (\bibinfo {year} {2007})},\ \bibinfo {note}
  {[,178(2007)]}\BibitemShut {NoStop}%
\bibitem [{\citenamefont {Nieves}\ \emph {et~al.}(2004)\citenamefont {Nieves},
  \citenamefont {Amaro},\ and\ \citenamefont {Valverde}}]{Nieves:2004wx}%
  \BibitemOpen
  \bibfield  {author} {\bibinfo {author} {\bibfnamefont {J.}~\bibnamefont
  {Nieves}}, \bibinfo {author} {\bibfnamefont {J.~E.}\ \bibnamefont {Amaro}},\
  and\ \bibinfo {author} {\bibfnamefont {M.}~\bibnamefont {Valverde}},\ }\href
  {https://doi.org/10.1103/PhysRevC.70.055503} {\bibfield  {journal} {\bibinfo
  {journal} {Phys. Rev. C}\ }\textbf {\bibinfo {volume} {70}},\ \bibinfo
  {pages} {055503} (\bibinfo {year} {2004})},\ \bibinfo {note} {[Erratum:
  \href{https://doi.org/10.1103/PhysRevC.72.019902}{Phys. Rev.C
  72,019902(2005)}]},\ \Eprint {https://arxiv.org/abs/nucl-th/0408005}
  {arXiv:nucl-th/0408005} \BibitemShut {NoStop}%
\bibitem [{\citenamefont {Gran}(2017)}]{Gran:2017psn}%
  \BibitemOpen
  \bibfield  {author} {\bibinfo {author} {\bibfnamefont {R.}~\bibnamefont
  {Gran}},\ }\href@noop {} {\  (\bibinfo {year} {2017})},\ \Eprint
  {https://arxiv.org/abs/1705.02932} {arXiv:1705.02932} \BibitemShut {NoStop}%
\bibitem [{\citenamefont {Nieves}\ \emph {et~al.}(2011)\citenamefont {Nieves},
  \citenamefont {Ruiz~Simo},\ and\ \citenamefont
  {Vicente~Vacas}}]{Nieves:2011pp}%
  \BibitemOpen
  \bibfield  {author} {\bibinfo {author} {\bibfnamefont {J.}~\bibnamefont
  {Nieves}}, \bibinfo {author} {\bibfnamefont {I.}~\bibnamefont {Ruiz~Simo}},\
  and\ \bibinfo {author} {\bibfnamefont {M.}~\bibnamefont {Vicente~Vacas}},\
  }\href {https://doi.org/10.1103/PhysRevC.83.045501} {\bibfield  {journal}
  {\bibinfo  {journal} {Phys.~Rev.~C}\ }\textbf {\bibinfo {volume} {83}},\
  \bibinfo {pages} {045501} (\bibinfo {year} {2011})},\ \Eprint
  {https://arxiv.org/abs/1102.2777} {arXiv:1102.2777 [hep-ph]} \BibitemShut
  {NoStop}%
\bibitem [{\citenamefont {Gran}\ \emph {et~al.}(2013)\citenamefont {Gran},
  \citenamefont {Nieves}, \citenamefont {Sanchez},\ and\ \citenamefont
  {Vicente~Vacas}}]{Gran:2013kda}%
  \BibitemOpen
  \bibfield  {author} {\bibinfo {author} {\bibfnamefont {R.}~\bibnamefont
  {Gran}}, \bibinfo {author} {\bibfnamefont {J.}~\bibnamefont {Nieves}},
  \bibinfo {author} {\bibfnamefont {F.}~\bibnamefont {Sanchez}},\ and\ \bibinfo
  {author} {\bibfnamefont {M.}~\bibnamefont {Vicente~Vacas}},\ }\href
  {https://doi.org/10.1103/PhysRevD.88.113007} {\bibfield  {journal} {\bibinfo
  {journal} {Phys.~Rev.~D}\ }\textbf {\bibinfo {volume} {88}},\ \bibinfo
  {pages} {113007} (\bibinfo {year} {2013})},\ \Eprint
  {https://arxiv.org/abs/1307.8105} {arXiv:1307.8105 [hep-ph]} \BibitemShut
  {NoStop}%
\bibitem [{\citenamefont {Schwehr}\ \emph {et~al.}(2016)\citenamefont
  {Schwehr}, \citenamefont {Cherdack},\ and\ \citenamefont
  {Gran}}]{Schwehr:2016pvn}%
  \BibitemOpen
  \bibfield  {author} {\bibinfo {author} {\bibfnamefont {J.}~\bibnamefont
  {Schwehr}}, \bibinfo {author} {\bibfnamefont {D.}~\bibnamefont {Cherdack}},\
  and\ \bibinfo {author} {\bibfnamefont {R.}~\bibnamefont {Gran}},\ }\href@noop
  {} {\  (\bibinfo {year} {2016})},\ \Eprint {https://arxiv.org/abs/1601.02038}
  {arXiv:1601.02038 [hep-ph]} \BibitemShut {NoStop}%
\bibitem [{\citenamefont {Rodrigues}\ \emph
  {et~al.}(2016{\natexlab{a}})\citenamefont {Rodrigues} \emph
  {et~al.}}]{Rodrigues:2015hik}%
  \BibitemOpen
  \bibfield  {author} {\bibinfo {author} {\bibfnamefont {P.~A.}\ \bibnamefont
  {Rodrigues}} \emph {et~al.} (\bibinfo {collaboration} {MINERvA}),\ }\href
  {https://doi.org/10.1103/PhysRevLett.116.071802} {\bibfield  {journal}
  {\bibinfo  {journal} {Phys. Rev. Lett.}\ }\textbf {\bibinfo {volume} {116}},\
  \bibinfo {pages} {071802} (\bibinfo {year} {2016}{\natexlab{a}})},\ \Eprint
  {https://arxiv.org/abs/1511.05944} {arXiv:1511.05944 [hep-ex]} \BibitemShut
  {NoStop}%
\bibitem [{\citenamefont {Ruterbories}\ \emph {et~al.}(2019)\citenamefont
  {Ruterbories} \emph {et~al.}}]{Ruterbories:2018gub}%
  \BibitemOpen
  \bibfield  {author} {\bibinfo {author} {\bibfnamefont {D.}~\bibnamefont
  {Ruterbories}} \emph {et~al.} (\bibinfo {collaboration} {MINERvA
  Collaboration}),\ }\href {https://doi.org/10.1103/PhysRevD.99.012004}
  {\bibfield  {journal} {\bibinfo  {journal} {Phys. Rev. D}\ }\textbf {\bibinfo
  {volume} {99}},\ \bibinfo {pages} {012004} (\bibinfo {year} {2019})},\
  \Eprint {https://arxiv.org/abs/1811.02774} {arXiv:1811.02774} \BibitemShut
  {NoStop}%
\bibitem [{\citenamefont {Rodrigues}\ \emph
  {et~al.}(2016{\natexlab{b}})\citenamefont {Rodrigues}, \citenamefont
  {Wilkinson},\ and\ \citenamefont {McFarland}}]{Rodrigues:2016xjj}%
  \BibitemOpen
  \bibfield  {author} {\bibinfo {author} {\bibfnamefont {P.}~\bibnamefont
  {Rodrigues}}, \bibinfo {author} {\bibfnamefont {C.}~\bibnamefont
  {Wilkinson}},\ and\ \bibinfo {author} {\bibfnamefont {K.}~\bibnamefont
  {McFarland}},\ }\href {https://doi.org/10.1140/epjc/s10052-016-4314-3}
  {\bibfield  {journal} {\bibinfo  {journal} {Eur. Phys. J. C}\ }\textbf
  {\bibinfo {volume} {76}},\ \bibinfo {pages} {474} (\bibinfo {year}
  {2016}{\natexlab{b}})},\ \Eprint {https://arxiv.org/abs/1601.01888}
  {arXiv:1601.01888 [hep-ex]} \BibitemShut {NoStop}%
\bibitem [{\citenamefont {Aliaga}\ \emph {et~al.}(2015)\citenamefont {Aliaga}
  \emph {et~al.}}]{Aliaga:2015aqe}%
  \BibitemOpen
  \bibfield  {author} {\bibinfo {author} {\bibfnamefont {L.}~\bibnamefont
  {Aliaga}} \emph {et~al.} (\bibinfo {collaboration} {MINERvA Collaboration}),\
  }\href {https://doi.org/10.1016/j.nima.2015.04.003} {\bibfield  {journal}
  {\bibinfo  {journal} {Nucl.~Instrum.~Meth.~A}\ }\textbf {\bibinfo {volume}
  {789}},\ \bibinfo {pages} {28} (\bibinfo {year} {2015})},\ \Eprint
  {https://arxiv.org/abs/1501.06431} {arXiv:1501.06431 [physics.ins-det]}
  \BibitemShut {NoStop}%
\bibitem [{\citenamefont {Mousseau}\ \emph {et~al.}(2016)\citenamefont
  {Mousseau} \emph {et~al.}}]{Mousseau}%
  \BibitemOpen
  \bibfield  {author} {\bibinfo {author} {\bibfnamefont {J.}~\bibnamefont
  {Mousseau}} \emph {et~al.} (\bibinfo {collaboration} {MINERvA}),\ }\href
  {https://doi.org/10.1103/PhysRevD.93.071101} {\bibfield  {journal} {\bibinfo
  {journal} {Phys. Rev. D}\ }\textbf {\bibinfo {volume} {93}},\ \bibinfo
  {pages} {071101} (\bibinfo {year} {2016})},\ \Eprint
  {https://arxiv.org/abs/1601.06313} {arXiv:1601.06313 [hep-ex]} \BibitemShut
  {NoStop}%
\bibitem [{\citenamefont {Michael}\ \emph {et~al.}(2008)\citenamefont
  {Michael}, \citenamefont {Adamson}, \citenamefont {Alexopoulos},
  \citenamefont {Allison}, \citenamefont {Alner}, \citenamefont {Anderson},
  \citenamefont {Andreopoulos}, \citenamefont {Andrews}, \citenamefont
  {Andrews}, \citenamefont {Arroyo},\ and\ \citenamefont
  {et~al.}}]{Michael_2008}%
  \BibitemOpen
  \bibfield  {author} {\bibinfo {author} {\bibfnamefont {D.}~\bibnamefont
  {Michael}}, \bibinfo {author} {\bibfnamefont {P.}~\bibnamefont {Adamson}},
  \bibinfo {author} {\bibfnamefont {T.}~\bibnamefont {Alexopoulos}}, \bibinfo
  {author} {\bibfnamefont {W.}~\bibnamefont {Allison}}, \bibinfo {author}
  {\bibfnamefont {G.}~\bibnamefont {Alner}}, \bibinfo {author} {\bibfnamefont
  {K.}~\bibnamefont {Anderson}}, \bibinfo {author} {\bibfnamefont
  {C.}~\bibnamefont {Andreopoulos}}, \bibinfo {author} {\bibfnamefont
  {M.}~\bibnamefont {Andrews}}, \bibinfo {author} {\bibfnamefont
  {R.}~\bibnamefont {Andrews}}, \bibinfo {author} {\bibfnamefont
  {C.}~\bibnamefont {Arroyo}},\ and\ \bibinfo {author} {\bibnamefont
  {et~al.}},\ }\href {https://doi.org/10.1016/j.nima.2008.08.003} {\bibfield
  {journal} {\bibinfo  {journal} {Nuclear Instruments and Methods in Physics
  Research Section A: Accelerators, Spectrometers, Detectors and Associated
  Equipment}\ }\textbf {\bibinfo {volume} {596}},\ \bibinfo {pages} {190–228}
  (\bibinfo {year} {2008})}\BibitemShut {NoStop}%
\bibitem [{\citenamefont {Adamson}\ \emph {et~al.}(2006)\citenamefont
  {Adamson}, \citenamefont {Crone}, \citenamefont {Jenner}, \citenamefont
  {Nichol}, \citenamefont {Saakyan}, \citenamefont {Smith}, \citenamefont
  {Thomas}, \citenamefont {Kordosky}, \citenamefont {Lang}, \citenamefont
  {Vahle}, \citenamefont {Belias}, \citenamefont {Nicholls}, \citenamefont
  {Pearce}, \citenamefont {Petyt}, \citenamefont {Barker}, \citenamefont
  {Cabrera}, \citenamefont {Hartnell}, \citenamefont {Miyagawa}, \citenamefont
  {Tagg}, \citenamefont {Weber}, \citenamefont {{Falk Harris}}, \citenamefont
  {Harris}, \citenamefont {Morse}, \citenamefont {Symes}, \citenamefont
  {Michael}, \citenamefont {Litchfield}, \citenamefont {Lee},\ and\
  \citenamefont {Boyd}}]{ADAMSON2006119}%
  \BibitemOpen
  \bibfield  {author} {\bibinfo {author} {\bibfnamefont {P.}~\bibnamefont
  {Adamson}}, \bibinfo {author} {\bibfnamefont {G.}~\bibnamefont {Crone}},
  \bibinfo {author} {\bibfnamefont {L.}~\bibnamefont {Jenner}}, \bibinfo
  {author} {\bibfnamefont {R.}~\bibnamefont {Nichol}}, \bibinfo {author}
  {\bibfnamefont {R.}~\bibnamefont {Saakyan}}, \bibinfo {author} {\bibfnamefont
  {C.}~\bibnamefont {Smith}}, \bibinfo {author} {\bibfnamefont
  {J.}~\bibnamefont {Thomas}}, \bibinfo {author} {\bibfnamefont
  {M.}~\bibnamefont {Kordosky}}, \bibinfo {author} {\bibfnamefont
  {K.}~\bibnamefont {Lang}}, \bibinfo {author} {\bibfnamefont {P.}~\bibnamefont
  {Vahle}}, \bibinfo {author} {\bibfnamefont {A.}~\bibnamefont {Belias}},
  \bibinfo {author} {\bibfnamefont {T.}~\bibnamefont {Nicholls}}, \bibinfo
  {author} {\bibfnamefont {G.}~\bibnamefont {Pearce}}, \bibinfo {author}
  {\bibfnamefont {D.}~\bibnamefont {Petyt}}, \bibinfo {author} {\bibfnamefont
  {M.}~\bibnamefont {Barker}}, \bibinfo {author} {\bibfnamefont
  {A.}~\bibnamefont {Cabrera}}, \bibinfo {author} {\bibfnamefont
  {J.}~\bibnamefont {Hartnell}}, \bibinfo {author} {\bibfnamefont
  {P.}~\bibnamefont {Miyagawa}}, \bibinfo {author} {\bibfnamefont
  {N.}~\bibnamefont {Tagg}}, \bibinfo {author} {\bibfnamefont {A.}~\bibnamefont
  {Weber}}, \bibinfo {author} {\bibfnamefont {E.}~\bibnamefont {{Falk
  Harris}}}, \bibinfo {author} {\bibfnamefont {P.}~\bibnamefont {Harris}},
  \bibinfo {author} {\bibfnamefont {R.}~\bibnamefont {Morse}}, \bibinfo
  {author} {\bibfnamefont {P.}~\bibnamefont {Symes}}, \bibinfo {author}
  {\bibfnamefont {D.}~\bibnamefont {Michael}}, \bibinfo {author} {\bibfnamefont
  {P.}~\bibnamefont {Litchfield}}, \bibinfo {author} {\bibfnamefont
  {R.}~\bibnamefont {Lee}},\ and\ \bibinfo {author} {\bibfnamefont
  {S.}~\bibnamefont {Boyd}},\ }\href
  {https://doi.org/https://doi.org/10.1016/j.nima.2005.10.072} {\bibfield
  {journal} {\bibinfo  {journal} {Nuclear Instruments and Methods in Physics
  Research Section A: Accelerators, Spectrometers, Detectors and Associated
  Equipment}\ }\textbf {\bibinfo {volume} {556}},\ \bibinfo {pages} {119}
  (\bibinfo {year} {2006})}\BibitemShut {NoStop}%
\bibitem [{\citenamefont {Diaz~Bautista}(2015)}]{DiazBautista:2015qyl}%
  \BibitemOpen
  \bibfield  {author} {\bibinfo {author} {\bibfnamefont {G.~A.}\ \bibnamefont
  {Diaz~Bautista}},\ }\emph {\bibinfo {title} {{Determinacion del error
  sistematico del momentum de muones producidos por interacciones
  neutrino-nucleon en el detector MINER$\nu$A}}},\ \href
  {https://doi.org/10.2172/1362136} {Master's thesis},\ \bibinfo  {school}
  {Lima, Pont. U. Catolica} (\bibinfo {year} {2015})\BibitemShut {NoStop}%
\bibitem [{\citenamefont {Valencia}\ \emph {et~al.}(2019)\citenamefont
  {Valencia} \emph {et~al.}}]{Valencia:2019mkf}%
  \BibitemOpen
  \bibfield  {author} {\bibinfo {author} {\bibfnamefont {E.}~\bibnamefont
  {Valencia}} \emph {et~al.} (\bibinfo {collaboration} {MINERvA}),\ }\href
  {https://doi.org/10.1103/PhysRevD.100.092001} {\bibfield  {journal} {\bibinfo
   {journal} {Phys. Rev. D}\ }\textbf {\bibinfo {volume} {100}},\ \bibinfo
  {pages} {092001} (\bibinfo {year} {2019})},\ \Eprint
  {https://arxiv.org/abs/1906.00111} {arXiv:1906.00111 [hep-ex]} \BibitemShut
  {NoStop}%
\end{thebibliography}%

\end{document}